\documentclass[sn-nature,Numbered]{sn-jnl}

\usepackage{subfiles}%
\usepackage{graphicx}%
\usepackage{multirow}%
\usepackage{amsmath,amssymb,amsfonts}%
\usepackage{amsthm}%
\usepackage{mathrsfs}%
\usepackage[title]{appendix}%
\usepackage{xcolor}%
\usepackage{soul}%
\usepackage{textcomp}%
\usepackage{siunitx}%
\usepackage{manyfoot}%
\usepackage{booktabs}%
\usepackage{algorithm}%
\usepackage{algorithmicx}%
\usepackage{algpseudocode}%
\usepackage{listings}%
\usepackage{floatpag}
\usepackage{etoc}
\usepackage{multibib}
\newcites{SI}{Supplementary References}
\newcites{Meth}{Methods References}

\begin{document}
\title[NeuralGCMs]{Neural general circulation models optimized to predict satellite-based precipitation observations}
\author*[1]{\fnm{Janni} \sur{Yuval}}
\email{janniyuval@google.com}
\equalcont{These authors contributed equally to this work.}
\author[1]{\fnm{Ian} \sur{Langmore}}
\equalcont{These authors contributed equally to this work.}
\author[1]{\fnm{Dmitrii} \sur{Kochkov}}
\equalcont{These authors contributed equally to this work.}
\author[1]{\fnm{Stephan} \sur{Hoyer}}
\equalcont{These authors contributed equally to this work.}
\affil[1]{\orgname{Google Research, Mountain View, CA}}


\abstract{Climate models struggle to accurately simulate precipitation, particularly extremes and the diurnal cycle. 
Here, we present a hybrid model that is trained directly on satellite-based precipitation observations. Our model runs at 2.8$^\circ$ resolution and is built on the differentiable NeuralGCM framework. The model demonstrates significant improvements over existing general circulation models, the ERA5 reanalysis, and a global cloud-resolving model in simulating precipitation. Our approach yields reduced biases, a more realistic precipitation distribution, improved representation of extremes, and a more accurate diurnal cycle. 
Furthermore, it outperforms the mid-range precipitation forecast of the ECMWF ensemble.
This advance paves the way for more reliable simulations of current climate and demonstrates how training on observations can be used to directly improve GCMs.}

\maketitle

\section*{Introduction \label{sec:intro}}


General Circulation Models (GCMs) are essential tools for understanding climate change and its impacts, yet they exhibit significant limitations in accurately representing precipitation, a key variable with profound societal implications.  These limitations manifest in both the spatial and temporal dimensions, and are especially severe when dealing with extreme precipitation. Spatially, biases in simulated precipitation patterns can be as large as the projected changes themselves \cite{PalmerStevens2019}, undermining confidence in model projections.  Temporally, GCMs struggle to accurately capture the diurnal cycle of precipitation \cite{dai2006precipitation,fiedler2020simulated, tang2021evaluating}, a critical factor influencing various hydrological processes, climate
variability, and weather forecasting. While observational data confirms a clear global trend in extreme precipitation \cite{o2015precipitation}, the limited observational record hinders the identification of robust regional changes in recent decades \cite{fischer2013robust}.  
The persistent difficulties models face in accurately simulating extreme precipitation restrict their utility for understanding regional trends in these high-impact events. There has been little improvement in this regard\cite{wehner2020characterization} from the 5th Coupled Model Intercomparison Project (CMIP5) to CMIP6.
 \cite{eyring2016overview}.
Given the critical societal implications of changes in precipitation \cite{trenberth2003changing,field2012managing}, there is an urgent need to improve the fidelity of precipitation simulations in GCMs.

The inaccurate representation of precipitation in current GCMs is largely attributed to deficiencies in deep convection parameterization schemes \cite{wilcox2007frequency}. To address this, three main approaches have been explored:
\begin{enumerate}
    \item Kilometer-scale global storm-resolving models \cite{stevens2019dyamond,slingo2022ambitious}, while promising, remain computationally prohibitive for long-term climate simulations and still exhibit their own limitations \cite{ma2022superior,feng2023mesoscale}.
     \item Purely machine learning-based atmospheric models have shown excellent results for short-term forecasting \cite{ravuri2021skilful,lam2022graphcast}. Recent work has even demonstrated the feasibility of running long-term simulations \cite{watt2023ace} and training models directly on satellite-based precipitation observations \cite{stock2024diffobs}. However, these models have yet to outperform traditional GCMs in terms of long-term climate statistics \cite{duncan2024application}.
    \item Hybrid models incorporating machine learning parameterizations can be run within a traditional GCM framework \cite{gentine2018could}.  So far, ML parameterizations in atmospheric models have heavily relied on data from high-fidelity simulations, such as convection-resolving models or super-parameterizations, rather than directly incorporating the vast amount of observational data  available from satellites, radiosondes, and ground-based instruments. This dependence arises from the difficulty of directly utilizing observational data to derive subgrid-scale tendencies or fluxes, which are the typical training targets for these parameterizations.  While there have been advancements in hybrid models \cite{rasp2018deep,yuval2020stable, yuval2021use}, challenges such as instabilities \cite{brenowitz2020interpreting}, climate drift \cite{brenowitz2019spatially}, and large biases \cite{kwa2023machine, han2023ensemble} are common.  Overall, under realistic conditions, hybrid models are still not competitive with existing GCMs for simulations of climate. Moreover, as long as ML parameterizations depend on high-fidelity simulations rather than observations, they will inevitably inherit the biases present in those simulations.
\end{enumerate}

Recently, hybrid modeling approach has been combined with differentiable dynamical core to enable end-to-end (i.e., ``online'') training. This led to the development of NeuralGCM\cite{kochkov2024neural}, a hybrid model trained on ERA5\cite{hersbach2020era5} data. NeuralGCM demonstrated the ability to run decadal simulations (albeit with occasional instabilities), exhibiting lower temperature biases in 40-year runs compared to AMIP-class models, along with a realistic seasonal cycle and state-of-the-art weather prediction skill. 
However, NeuralGCM trained solely on ERA5 data inherits all of the associated limitations, such as deficiencies in reproducing extreme precipitation events \cite{lavers2022evaluation} and the diurnal cycle of precipitation\cite{tang2020have}.

Building upon the NeuralGCM differentiable framework, we develop a hybrid model trained directly on satellite-based precipitation observations. By leveraging observational data, we demonstrate significant improvements in precipitation simulation both for weather forecasting and on simulations of climate compared to CMIP6 models, ERA5 reanalysis, and a Global Cloud-Resolving Model (GCRM).



\section*{Training a hybrid model from observations}

\begin{figure*}
\begin{center}
\makebox[\textwidth]{\colorbox{white}{\includegraphics[width=0.65\paperwidth]{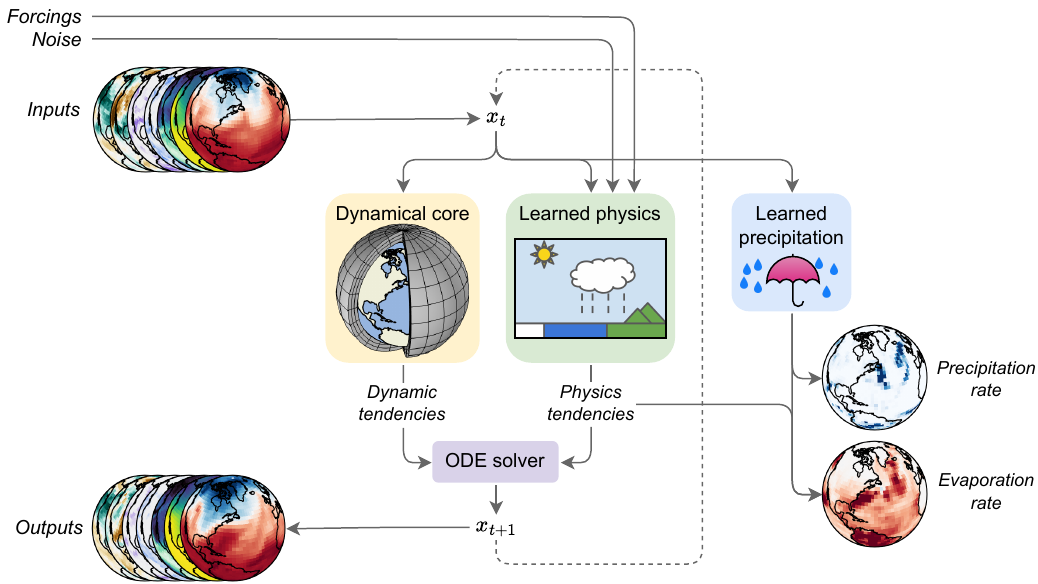}}}
\end{center}
\caption{
Overall model structure. Inputs are encoded into the model state $x_{\rm{t}}$.This state is fed into the dynamical core and the learned precipitation module. Along with forcings and noise, the state is also used as input to the learned physics module. The dynamical core and learned physics module produce tendencies (rates of change) for an implicit-explicit ordinary differential equation (ODE) solver, which advances the state in time to $x_{\rm{t+1}}$. The precipitation module predicts the precipitation rate and, by enforcing water column conservation (Eq.~\ref{eq:p_minus_e}), diagnoses the evaporation rate. The new model state can then be used for the next time step or decoded to produce outputs.
}\label{fig:model_structure}
\end{figure*}

In essence, NeuralGCM comprises two core components (Fig.~\ref{fig:model_structure}): (a) a differentiable dynamical core, and (b) a learned physics module (i.e., a neural network parameterization). This architecture results in a fully differentiable model, facilitating end-to-end (online) training\cite{kochkov2024neural}.
Within a differentiable model, optimization of model parameters requires only that the loss can be evaluated based on the ground truth and quantities accessible from the model predictions.
This allows learning via minimization of a loss comparing observations to model output. While any observational dataset could theoretically be employed, we elected to focus on precipitation, as it is a key variable that both models and reanalysis data struggle to simulate accurately.

The process of training NeuralGCM models with satellite-based precipitation observations follows the stochastic training approach of \cite{kochkov2024neural}, minimizing the continuous ranked probability score (CRPS)\cite{Gneiting2007ProperScoring} between predicted weather trajectories and the ground truth. We gradually increase the rollout length of these trajectories from 6 hours to 5 days. The trajectories are sampled from ERA5 for atmospheric variables and evaporation, and from the Integrated Multi-satellitE Retrievals for Global Precipitation Measurement (IMERG) V07 ``final'' dataset \cite{huffman2020integrated} for precipitation.  However, to incorporate precipitation optimization while preserving physical consistency and stability, we introduced several key modifications to the NeuralGCM models, as detailed below.

\subsection*{From water budget to precipitation and evaporation}
The original version of NeuralGCM\cite{kochkov2024neural} does not explicitly represent precipitation and evaporation. Instead, only the net precipitation minus evaporation ($P-E$) is diagnosed using the column water budget (Eq.~\ref{eq:p_minus_e}; Methods). Our objective now is to incorporate a precipitation variable in a manner consistent with the water budget, ensuring plausible values for both evaporation and precipitation. To achieve this, we introduce a neural network that predicts precipitation rate from the atmospheric column state (Eq.~S3) and diagnose evaporation by enforcing the column water budget (Eq.~S4).
In the supplementary information, we also present an alternative NeuralGCM configuration, referred to as NeuralGCM-evap, which utilizes a neural network to predict evaporation rate from surface variables, with precipitation diagnosed by enforcing the column water budget (Eq.~S2). 
We find that NeuralGCM-evap is in many aspects superior to the presented model, but one significant disadvantage is that it does not enforce non-negative precipitation.


We optimize for temperature, geopotential, zonal and meridional wind, specific humidity, specific water/ice cloud variables, hourly evaporation rate (from ERA5), and 6-hour accumulated precipitation (from IMERG). Optimization occurs every six hours, and the model we train has a $2.8^{\circ}$ grid spacing. 




Simultaneously optimizing NeuralGCM for both IMERG precipitation and ERA5 data presents inherent challenges. This arises from the inconsistency between ERA5 precipitation (and its associated moisture budget) and IMERG precipitation, where ERA5 often exhibits substantial deviations from IMERG, even when both datasets are coarse-grained to $2.8^{\circ}$ resolution (Fig.~S1; see Methods for a description of how we coarse-grain IMERG data in time). Consequently, using both ERA5 water variables (i.e., specific humidity, cloud variables, and evaporation rate) and IMERG precipitation for optimization introduces conflicting objectives.
In the supplementary information and in Fig.~S2, we demonstrate the potential advantages of incorporating physically consistent representations of precipitation and evaporation within NeuralGCM (rather than predicting precipitation without considering the column water budget).

Given our primary goal of enhancing precipitation representation, we have opted to slightly relax the constraint on accurately simulating specific humidity from ERA5 by reducing the corresponding loss weight (see Methods for how we determine loss weights), while still emphasizing both precipitation and evaporation. 
This relaxation is supported by the fact that ERA5 specific humidity exhibits non-negligible differences compared to observations\cite{johnston2021evaluation, kruger2022vertical}, justifying a greater tolerance for deviations from ERA5 in our model.
In the supplementary information, we also describe several additional modifications to NeuralGCM which enhance its stability, as well as limitations of our model.

\section*{Results \label{sec:results}}
We train a NeuralGCM model using data from 2001-2018. 
For both weather forecast results and climate results we regrid all datasets to a $2.8^\circ$ Gaussian grid using conservative regridding.
We then evaluate the skill of the NeuralGCM model for both weather forecasting and long integrations for climate simulations.

We consider both IMERG and the Global Precipitation Climatology Project \cite{huffman2023new} (GPCP; a dataset not used in training) as ground truth for precipitation. These datasets were chosen due to their extensive use and established reliability as benchmarks for precipitation in climate science\cite{hersbach2020era5,nogueira2020inter,ma2022superior,feng2023mesoscale}, providing robust standards for evaluating the performance of NeuralGCM.

Extensive literature comparing precipitation datasets demonstrates that IMERG and GPCP generally outperform reanalysis data, particularly ERA5, across various metrics and timescales. These include evaluations of diurnal cycles \cite{watters2021diurnal}, extreme precipitation \cite{jiang2023evaluation}, and monthly or longer accumulations compared to gauge measurements \cite{jiang2023evaluation, xin2022evaluation, wu2023statistical}. However, discrepancies exist in assessments of daily or shorter timescales, with some studies favoring IMERG over ERA5 in certain regions \cite{jiang2023evaluation, aryastana2023quantitative} while others suggest ERA5 may be more accurate in specific locations \cite{xin2022evaluation}. 

It is important to acknowledge that all precipitation products have inherent limitations \cite{sun2018review}. Specifically, IMERG's calibration process can lead to underestimation of light precipitation and overestimation of heavy precipitation \cite{pradhan2022review}. However, utilizing coarser spatiotemporal scales, as in this study, generally improves agreement between precipitation products \cite{herold2017large}, particularly between the NOAA Multi-Radar Multi-Sensor system \cite{zhang2016multi} and IMERG \cite{guilloteau2020multiscale} and between IMERG and gauge measurements at sub-daily timescales \cite{zhou2023evaluation}.

\subsection*{Medium-range precipitation forecasting}
For weather forecasting, we use the WeatherBench2\cite{rasp2024weatherbench} code to evaluate an ensemble of 50 NeuralGCM forecasts for 732 initial conditions at noon and midnight UTC spanning the year 2020, which was held out from the training data. We compare NeuralGCM results to those from the 50-member ECMWF ensemble (ENS) and probabilistic climatology (Methods).

\begin{figure*}
\begin{center}
\makebox[\textwidth]{\colorbox{white}{\includegraphics[width=0.8\paperwidth]{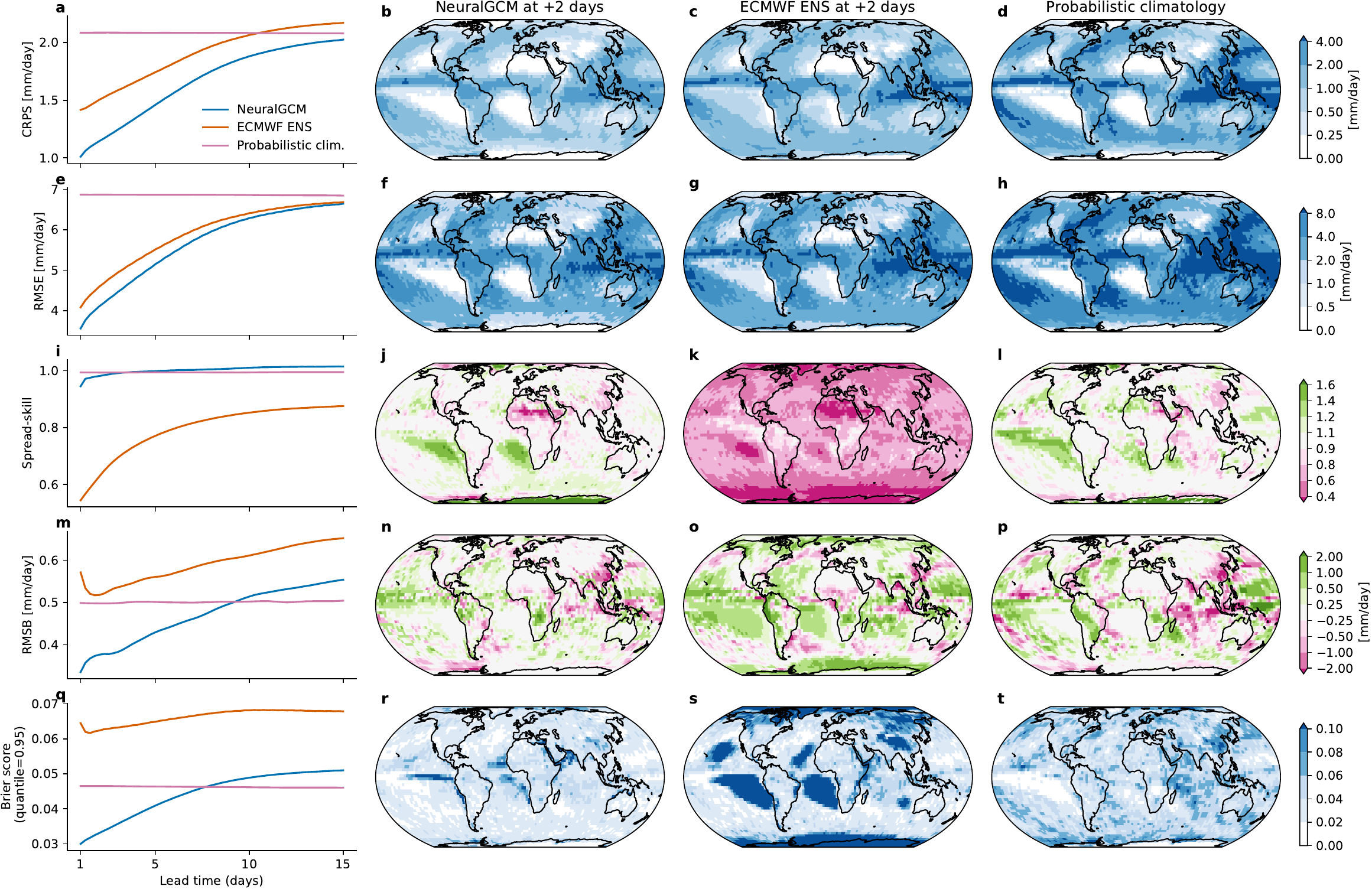}}}
\end{center}
\caption{
Precipitation forecasting accuracy scores for 24-hour accumulated precipitation, evaluated against IMERG. Area-weighted mean, calculated over all longitudes and latitudes between $-60^\circ$ to $60^\circ$ for: (a) Continuous Ranked Probability Score (CRPS). (e) Ensemble mean root-mean-square error (RMSE). (i) Spread-skill ratio. (m) Root-mean-square bias (RMSB). (q) Brier score (0.95 quantile). Comparisons are shown for NeuralGCM, the ECMWF ensemble, and probabilistic climatology (see Methods). Spatial distributions of (b, c, d) CRPS, (f, g, h) RMSE, (j, k, l) spread-skill ratio, (n, o, p) RMSB, and (r, s, t) Brier score (0.95 quantile) for NeuralGCM, the ECMWF ensemble, and probabilistic climatology on the second day of forecasting.
}\label{fig:evals_24h_spatial}
\end{figure*}

We find that NeuralGCM at $2.8^{\circ}$ significantly outperforms ENS in precipitation prediction across all 15 forecast days in terms of continuous ranked probability score (CRPS), ensemble-mean root-mean-square bias (RMSB), spread-skill ratio, and Brier score (0.95 quantile; see Methods). These results holds for both 24-hour (Fig.~\ref{fig:evals_24h_spatial}) and 6-hour accumulated precipitation (Fig.~S3) when evaluated against IMERG, including when evaluations are restricted to land regions (Fig.~S4).  NeuralGCM also outperforms ENS when evaluated against 24-hour accumulated precipitation from GPCP (Fig.~S5). 
NeuralGCM shows higher skill than probabilistic climatology for CRPS and RMSE for 15 days but has a larger RMSB and Brier score after 9 and 7 days, respectively.
NeuralGCM provides reasonable predictions for other variables but underperforms ENS, as expected given the low resolution of the current NeuralGCM configuration (Fig.~S6).
Sub 6-hour precipitation accumulations in NeuralGCM (but not NeuralGCM-evap) also show unrealistic oscillations in intensity, particularly during the first day of forecasting (Fig.~S7).

\subsection*{Precipitation in climate simulations}
To test the skill of NeuralGCM in simulating precipitation for climate simulations, we conducted 20-year simulations using 37 initial conditions spaced every 10 days throughout the year 2001.  For these simulations, we prescribed historical sea surface temperatures (SSTs) and sea ice concentrations. All 37 initial conditions remained stable for the full 20-year duration for the precipitation model presented in the main text.

We compared various aspects of precipitation in our model to CMIP6 models, ERA5 reanalysis data, and GFDL's X-SHiELD global cloud-resolving model\cite{cheng2022impact}. These included mean precipitation (Fig.~\ref{fig:mean_precip}), extreme precipitation and precipitation rate (Fig.~\ref{fig:rx1day_precip_and_distribution}), diurnal cycle (Fig.~\ref{fig:diurnal_cycle}), and the time-space spectrum (Fig.~S8). 

To investigate the sensitivity of extreme precipitation to global mean temperature changes within NeuralGCM, we conducted an extended analysis comprised of 732 ensemble runs of 22 years each. All but one of these runs remained stable for the full simulation period. The results of this analysis are presented in the supplementary information and illustrated in Fig.~S30.


Unless stated otherwise, we always use the NeuralGCM simulation initialized on December 27, 2001, for comparison. When comparing against X-SHiELD, we use the available dates in X-SHiELD (January 18, 2020, to January 17, 2021) for all relevant models. When comparing against AMIP or historical runs, we compare the years 2002-2014 (2014 is the last year which is available for AMIP runs).

\begin{figure*}
\begin{center}
\makebox[\textwidth]{\colorbox{white}{\includegraphics[width=0.8\paperwidth]{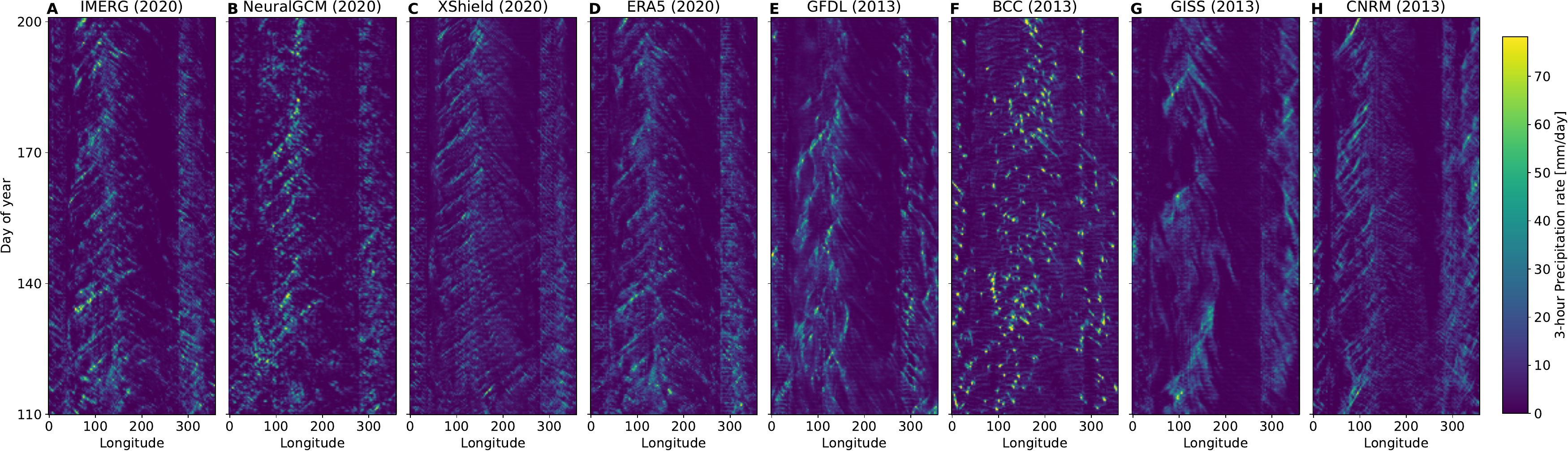}}}
\end{center}
\caption{
Hovmoller tropical precipitation diagram for different models.
Precipitation is averaged between latitudes $-5^{\circ}$ and $5^{\circ}$.
IMERG, NeuralGCM, X-SHiELD, and ERA5 data are shown for 91 days starting on April 20, 2020. CMIP model are shown for historical runs for 91 days starting on April 20, 2013.
NeuralGCM run shown was initialized on December 27 2001. All models were coarse-grained to 2.8$^{\circ}$ before plotting.
}\label{fig:hovmoller}
\end{figure*}

To visually demonstrate the differences between models, we show a Hovmöller diagram\cite{hovmoller1949trough} of 6 months of tropical precipitation from IMERG, NeuralGCM, ERA5, X-SHiELD, and several models from CMIP6 historical runs (Fig. \ref{fig:hovmoller}). Qualitatively, NeuralGCM exhibits the most similar structure to IMERG, both in terms of spatial structure and amplitude. All other models show substantial differences in both precipitation magnitude and spatiotemporal structure. ERA5, due to its assimilation process, has a very similar spatiotemporal structure to IMERG but fails to capture heavy precipitation rates. In the following analysis, we quantify further aspects of the simulated precipitation and show that NeuralGCM is not only visually compelling, but also statistically superior to the other models.


\subsection*{Mean precipitation}
\begin{figure*}
\begin{center}
\makebox[\textwidth]{\colorbox{white}{\includegraphics[width=0.65\paperwidth]{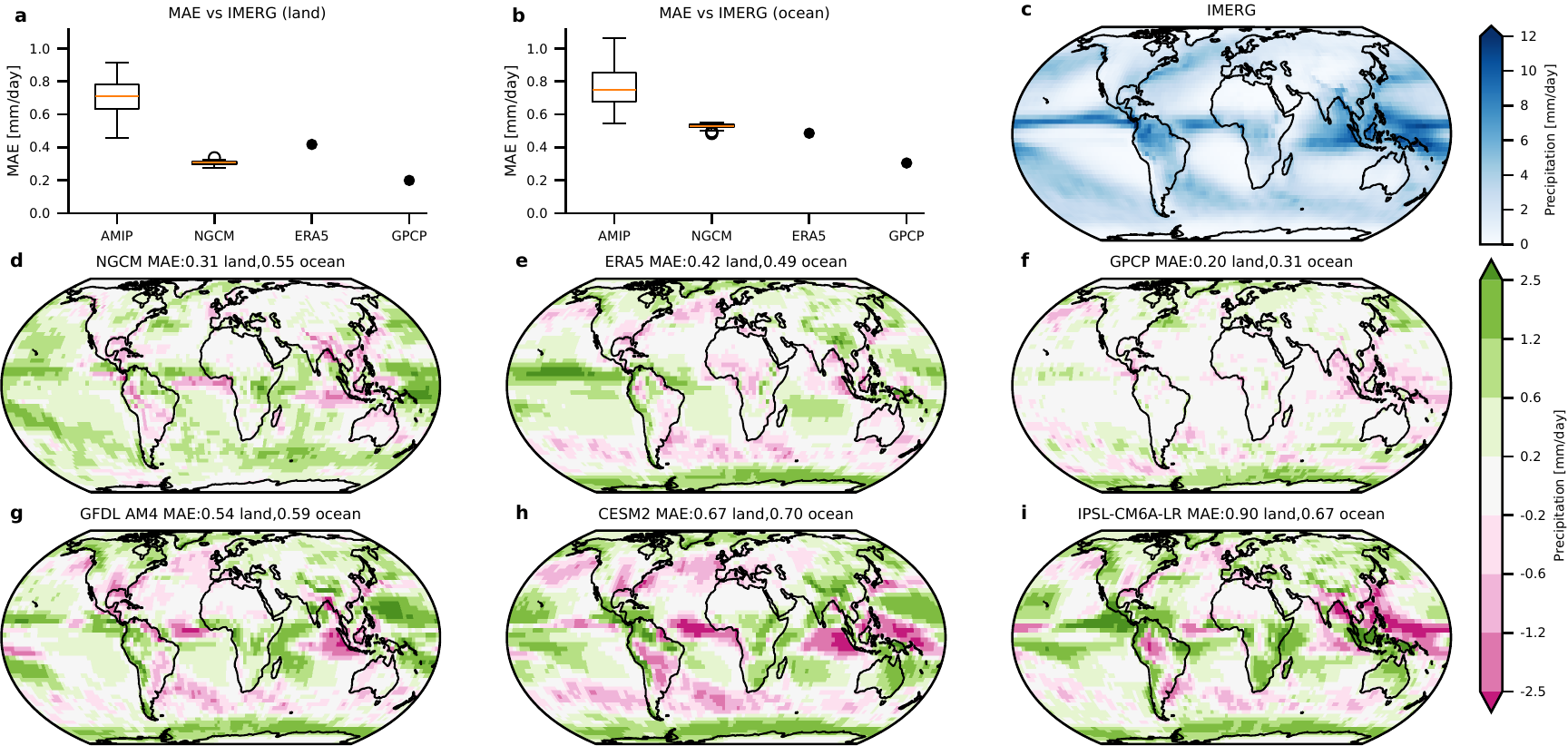}}}
\end{center}
\caption{
Bias in mean precipitation averaged over 2002–2014. (a, b) Box plots showing the mean absolute error (MAE) relative to IMERG for 37 NeuralGCM runs (initialized during 2001), 37 CMIP6 AMIP experiments (model details in Methods), ERA5, and GPCP\cite{huffman2023new} over (a) land and (b) ocean. In the box plots, the red line indicates the median; the box delineates the interquartile range (IQR); whiskers extend to 1.5 × IQR; and outliers are shown as dots. (c) IMERG mean precipitation averaged over 2002–2014. (d–i) Bias in mean precipitation from NeuralGCM, ERA5, GPCP, and three CMIP6 AMIP experiments. Global MAE (in mm/day) is shown for land and ocean regions.
}\label{fig:mean_precip}
\end{figure*}

Figure \ref{fig:mean_precip} shows the mean precipitation averaged over 2002-2014 for NeuralGCM, ERA5, and 37 CMIP6 AMIP experiments, compared to IMERG observations. Analysis of 37 NeuralGCM runs reveals a global mean absolute error (MAE) of 0.45 mm/day (0.30 mm/day over land, 0.52 mm/day over ocean), compared to 0.74 mm/day (0.76 mm/day over land, 0.70 mm/day over ocean) for 37 AMIP runs, representing a 40\% error reduction. Notably, NeuralGCM achieves a similar MAE to ERA5, which is particularly impressive given that NeuralGCM was run freely (forced by SST and sea-ice extent), while ERA5 assimilated observations every 12 hours.
This superior performance of NeuralGCM compared to AMIP simulations persists across individual seasons (Figs.~S11, S12, S13, S14) and when evaluated against GPCP data, which NeuralGCM was not trained on (Fig.~S15).

\subsection*{Precipitation extremes and precipitation rate distribution}
We examine the model's ability to reproduce the frequency distribution of 24-hourly precipitation rates, a challenging aspect of precipitation simulation that is sensitive to the choice of convection scheme \cite{wilcox2007frequency} and often poorly represented in CMIP-class models \cite{norris2021evaluation}. 
We estimate frequency distribution using 50 equally spaced bins in the logarithm of the precipitation rate, with lowest bin starting at 0.03 mm/day and the largest bin at 240 mm/day. We normalize the distribution such that it integrates to one when considering the whole distributions (including rates below 0.03 mm/day). We compare the frequency distributions of NeuralGCM, ERA5, and a single CMIP6 model (IPSL-CM6A-LR) to that of IMERG. We show results for IPSL-CM6A-LR as a representative example of a CMIP6 model to maintain clarity in the figure, but we acknowledge that different models have different distributions.

\begin{figure*}
\begin{center}
\makebox[\textwidth]{\colorbox{white}{\includegraphics[width=0.65\paperwidth]{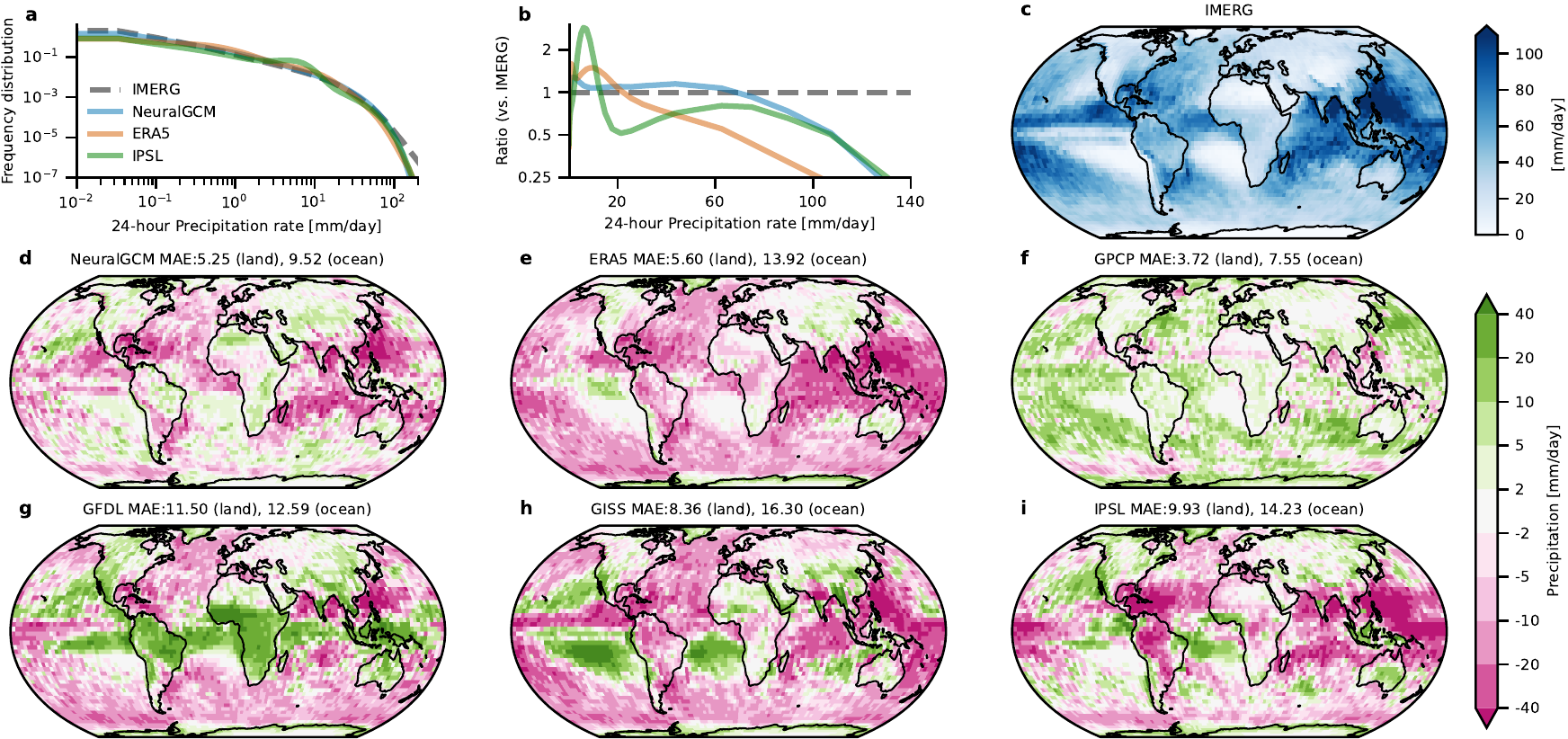}}}
\end{center}
\caption{
Tropical precipitation rate distribution and annual maximum daily precipitation (Rx1day) averaged over 2002–2014. (a) Frequency distributions of 24-hourly precipitation rate for IMERG\cite{huffman2020integrated}, NeuralGCM, ERA5, and IPSL-CM6A-LR (historical run) in the tropics (latitudes  -20$^\circ$ to 20$^\circ$). (b) Relative distribution normalized by the IMERG value. (c) IMERG Rx1day calculated over 2002-2014. (d–i) Bias in Rx1day for NeuralGCM, ERA5, GPCP\cite{huffman2023new}, and various CMIP6 historical simulations, relative to IMERG.  Global mean absolute error (MAE) relative to IMERG is shown for land and ocean regions (in mm/day). The NeuralGCM simulation was initialized on December 27, 2001. All models were coarsened to a $2.8^{\circ}$ resolution.
}\label{fig:rx1day_precip_and_distribution}
\end{figure*}

We find that the NeuralGCM frequency distribution of precipitation rates in the tropics is closer to the distribution from IMERG for both light and extreme precipitation than that of ERA5, IPSL-CM6A-LR (Fig.~\ref{fig:rx1day_precip_and_distribution}a,b) and X-SHiELD (Fig.~S9)). 
However, NeuralGCM underestimates the most extreme precipitation rates, which is partly due to their nature as grid-scale events (see also Fig.~S1). When the models are further regridded to a 5.6$^{\circ}$ resolution, NeuralGCM more closely follows the extreme precipitation rate occurrences in IMERG (Fig.~S10). 


To assess the ability of NeuralGCM to simulate the spatial patterns of extreme precipitation, we use the annual maximum daily precipitation at each grid point (often referred to as the Rx1day index; Fig.~\ref{fig:rx1day_precip_and_distribution}). We find that NeuralGCM represents Rx1day more accurately than ERA5 and the three CMIP6 models included in this comparison, 38--54\% reduction  in mean absolute error (MAE) over land compared to the CMIP6 models. NeuralGCM's MAE is only 25\% larger than GPCP's MAE, which as another observation-based product provides an estimate of observational uncertainty in IMERG.
Furthermore, NeuralGCM outperforms ERA5 and CMIP6 simulations when evaluated for the percent deviation from IMERG Rx1day (Fig.~S23 highlights regions outside the tropics). We find similar conclusions when studying the 99.9th percentile (Fig.~S22).



\subsection*{Diurnal cycle of precipitation}
Following previous studies\cite{dai2001global, tang2021evaluating}, we characterize the diurnal cycle of precipitation by the local solar time (LST) of maximum precipitation and the amplitude of the diurnal and semi-diurnal harmonics (see Methods). Similar to previous work\cite{tang2021evaluating}, we focus on the warm season in both hemispheres, where the diurnal cycle is more pronounced. 

\begin{figure*}
\begin{center}
\makebox[\textwidth]{\colorbox{white}{\includegraphics[width=0.65\paperwidth]{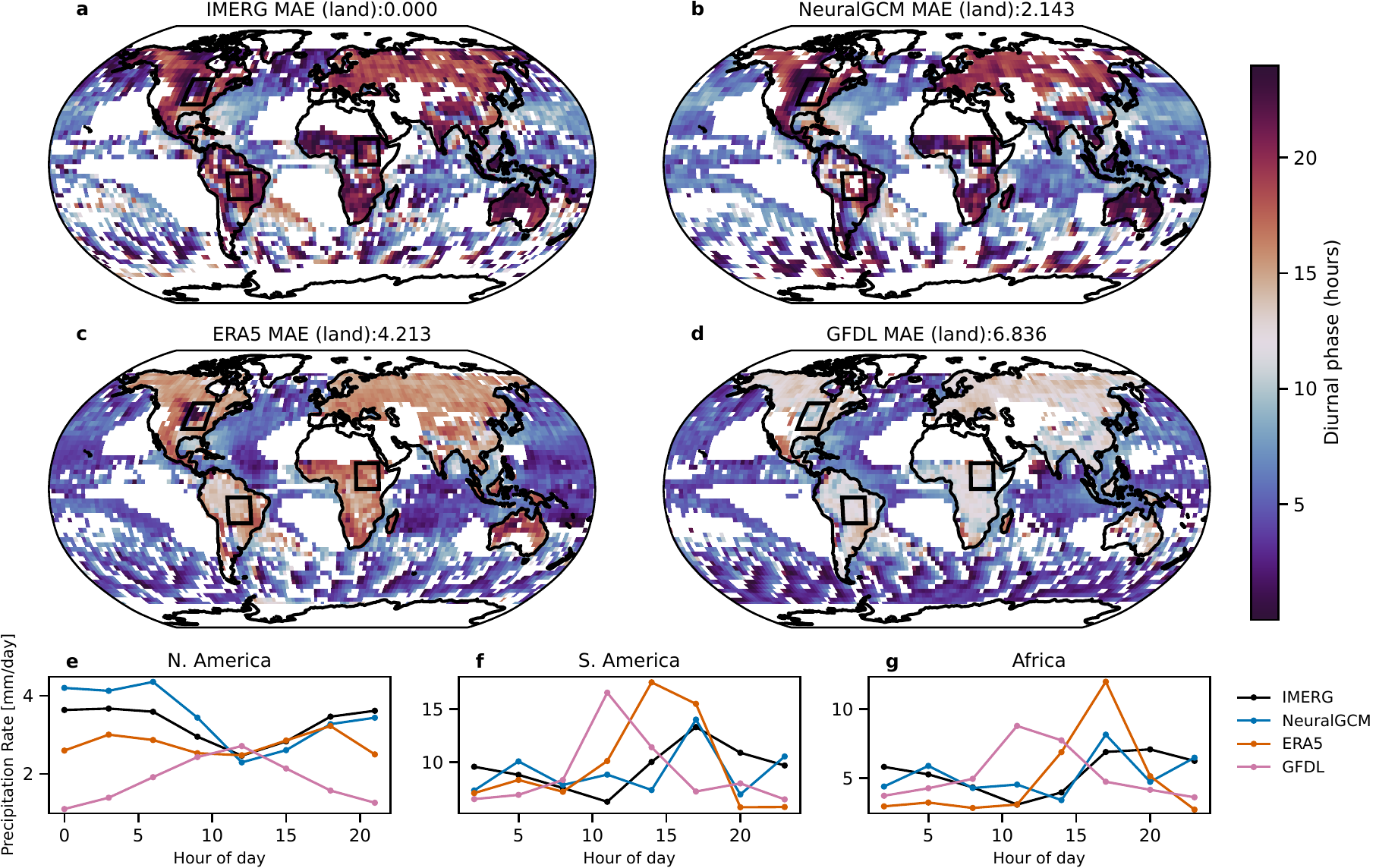}}}
\end{center}
\caption{
Diurnal Cycle of Summertime Precipitation (2002-2014)
(a-d) Local solar time (LST) of maximum precipitation during summertime (July in the Northern Hemisphere and January in the Southern Hemisphere) derived from the diurnal harmonic for (a) IMERG, (b) NeuralGCM, (c) ERA5 reanalysis, and (d) GFDL AMIP simulation. Regions where either the monthly mean precipitation is less than 0.75 mm/day or the diurnal amplitude ratio (amplitude normalized by mean precipitation) is less than 0.1 are masked in white.
Mean absolute error is calculated only above land.
(e-g)  Summertime diurnal cycle of precipitation (2002-2014) over subregions of (e) N. America, (f) S. America, and (g) Africa (indicated by rectangles in the maps).
}\label{fig:diurnal_cycle}
\end{figure*}

Fig.~\ref{fig:diurnal_cycle} demonstrates that NeuralGCM more accurately captures the timing of peak diurnal precipitation compared to ERA5 and GFDL AMIP run, both of which exhibit an early bias which has been an issue in models for decades\cite{trenberth2003changing}, particularly over land. 
NeuralGCM also exhibits a lower MAE for diurnal and semi-diurnal amplitude, as well as semi-diurnal phase (Figs.~S16, S17, S18).
However, as noted previously, the diurnal cycle in NeuralGCM exhibits unrealistic features, with certain times of day experiencing significantly more precipitation than others (Figs. \ref{fig:diurnal_cycle}e-g, S7), likely due to the model being optimized for 6-hourly precipitation accumulation. These unrealistic diurnal features are not present in NeuralGCM-evap (Figs. S7, S27).


\section*{Discussion}

By harnessing a differentiable dynamical core and a neural network parameterization, NeuralGCM can be trained jointly on ERA5 and observational products, providing a compelling example of how observational knowledge can enhance the fidelity of atmospheric simulations.
When trained on satellite-based precipitation observations, NeuralGCM remains stable for decadal simulations and substantially surpasses traditional GCMs and ERA5 in accurately simulating key aspects of precipitation, including its mean state, extremes, and the diurnal cycle.

While this study employed a neural network to parameterize all processes unresolved by the dynamical core, future work could explore coupling our differentiable dynamical core with a traditional parameterization suite and optimizing its free parameters. This approach offers the potential to further refine existing parameterizations by leveraging observational data. Moreover, it could reveal inherent limitations in the structure of current parameterizations, guiding the development of more accurate and physically consistent representations of unresolved processes.

Although our model has a lower resolution than typical models used for weather forecasts of precipitation, which limits its immediate practical applications, it demonstrates that a low-resolution hybrid model can substantially outperform ECMWF's ensemble prediction system in precipitation prediction. This suggests that further improvements in resolution, achieved through statistical downscaling or a higher-resolution model, could yield substantial gains compared to ECMWF's model.

Our work retains some noteworthy limitations. While the presented NeuralGCM is much more stable than prior models \cite{kochkov2024neural}, the stable model was still obtained by training several models with varying random seeds and choosing the most stable one. Further research is needed to understand and address the factors that influence model stability.
Finally, developing effective strategies for learning from potentially conflicting datasets is crucial. In this study, we encountered inconsistencies between ERA5 and IMERG, necessitating careful tuning of the loss function. Ideally, future research will also prioritize the development of unified datasets to provide a single, consistent ground truth for model training, thereby avoiding the need for ad hoc adjustments. 

\subsection*{Code availability}
NeuralGCM code base is publicly available on GitHub at  \url{https://github.com/neuralgcm/neuralgcm}.

\subsection*{Data availability}
3-hourly outputs from 20-year simulations of the NeuralGCM precipitation model are available via Google Cloud Storage in Zarr format at \path{gs://neuralgcm/amip_runs/v1_precip_stochastic_2_8_deg/2001-to-2021_128x64_gauss_37-level_stride3h.zarr}. NeuralGCM model checkpoint can be found at \path{gs://neuralgcm/models/v1_precip/stochastic_precip_2_8_deg.pkl}. NeuralGCM-evap model checkpoint can be found at \path{gs://neuralgcm/models/v1_precip/stochastic_evap_2_8_deg.pkl}.
The GitHub repository provides examples of how to use NeuralGCM checkpoints for simulations.

IMERG \cite{huffman2020integrated} data were downloaded from \path{ftps://arthurhouftps.pps.eosdis.nasa.gov/gpmdata}. 
GPCP \cite{huffman2023new} data were downloaded from \url{https://disc.gsfc.nasa.gov/datasets/GPCPDAY_3.2/summary}.
ERA5 \cite{hersbach2020era5} was originally downloaded from \url{https://cds.climate.copernicus.eu/} and is available via Google Cloud Storage in Zarr format at \path{gs://gcp-public-data-arco-era5/ar/full_37-1h-0p25deg-chunk-1.zarr-v3}.
CMIP6 data available at \url{https://catalog.pangeo.io/browse/master/climate/}.

\section*{Methods \label{sec:methods}}

\subsection*{Neural Networks}

\subsubsection*{Neural network for predicting tendencies}
NeuralGCM's neural network (NN) parameterization for predicting tendencies adopts the single-column approach common in GCMs, where information from a single atmospheric column is used to predict the impact of unresolved processes within that column. A fully connected neural network with residual connections is employed for this prediction, with the network weights shared across all columns.

A full description of the NN parameterization (i.e., the NN that predicts tendencies), its architecture, features, and parameters, is detailed in the supplementary material of \cite{kochkov2024neural}. The main difference in this work compared to our previous paper is that the parameterization also predicts tendencies for log surface pressure, which significantly improved stability in multi-year simulations.

\subsubsection*{Neural network for predicting precipitation}
Here, we employ an additional single-column network to predict precipitation (at 1-hour intervals), but with different parameters and inputs. Overall, the precipitation network is similar to the parameterization network, but it is much smaller. The features and architecture of the precipitation NN are described below. 
 
The core input features to the neural network include the vertical profiles of zonal and meridional wind, temperature anomalies, specific humidity, specific cloud ice water content, and specific cloud liquid water content. Unlike in the NN parameterization for predicting tendencies, we do not include the spatial derivatives of these fields as inputs.
We also include orography (along with its spatial gradients), a land-sea mask, and an 8-dimensional location-specific embedding vector for each horizontal grid point. This embedding vector aims to represent static, location-specific information related to precipitation (e.g., subgrid orography). It is initialized with random values and optimized during training.

Additionally, we use a surface embedding network that receives surface-related inputs, specifically sea surface temperature (SST) and sea ice concentration. Over land and ice where SST is not available, we include the lowest model level temperature and specific humidity. (Full details are provided in\cite{kochkov2024neural}.)

It is important to note that the learned embedding vector and the surface embedding network for the precipitation NN have different parameters than those used in the NN parameterization. All features are normalized to have an approximate zero mean and unit variance to improve training dynamics, as described in\cite{kochkov2024neural}.

Similar to the NN parameterization for predicting tendencies, we use a fully connected neural network with residual connections\cite{kochkov2024neural}. However, this network predicts only precipitation. We employ an Encode-Process-Decode (EPD) architecture\cite{battaglia2018relational} with 3 fully connected MLP blocks in the ``Process'' component (compared to 5 blocks in the NN parameterization for predicting tendencies).

All input features are concatenated and passed to the ``Encode'' layer, a linear layer that maps the input features to a latent vector of size 64 (compared to 384 in the NN parameterization for predicting tendencies). Each ``Process'' block utilizes a 3-layer MLP with 64 hidden units (compared to 384 for the NN parameterization for predicting tendencies) to update the latent vector. Finally, a linear ``Decode'' layer maps the latent vector of size 64 (384 in the NN parameterization for predicting tendencies) to the hourly precipitation rate. A ReLU activation function is then applied to ensure non-negativity of the predicted precipitation.

\subsection*{Variable re-scaling for losses}

To balance the contributions of different variables to the loss function, we rescaled the losses following a similar approach to that in our previous work \cite{kochkov2024neural}. Specifically, we divided each atmospheric variable by the standard deviation of its temporal difference over 24 hours and applied a time-dependent rescaling function \cite{kochkov2024neural}. However, we reduced the scaling factor for specific humidity by a factor of 100 to discourage the model from closely following ERA5 estimates of specific humidity. This adjustment allowed us to achieve precipitation values closer to IMERG. The scaling factors for precipitation and evaporation were determined empirically to ensure that these variables contributed approximately 10\% and 20\%, respectively, to the total loss, while specific humidity contributed only 3\%.

\subsection*{Water budget in NeuralGCM model}
Precipitation minus evaporation is diagnosed by integrating the moisture budget tendencies from the NN parameterization for tendencies:
\begin{equation}\label{eq:p_minus_e}
    P-E = \frac{1}{g}\int_{0}^{1} \sum_{\rm{i}} \left(\frac{dq}{dt}\right)_{\rm{i}}^{\rm{NN_{\rm{tend}}}} p_{\rm{s}} d\sigma
\end{equation}
where $p_{\rm{s}}$ is the surface pressure, and $\sum_{\rm{i}} (\frac{dq}{dt})_{\rm{i}}^{\rm{NN_{\rm{tend}}}}$ is the sum of the water species (i.e., specific humidity $q$, specific cloud ice $q_{c_i}$ and specific liquid cloud water content $q_{c_l}$) tendencies predicted by the neural network.

\subsection*{Diurnal cycle of precipitation}
Following previous studies\cite{dai2001global, tang2021evaluating}, we apply Fourier analysis to the diurnal time series of precipitation.  (The data is first grouped by hour and averaged.) The 3-hourly precipitation time series,  $P(t)$, $t\in\{0\ldots,23\}$,  is then represented as:
\begin{equation}
P(t) = S_0 + S_1(t) + S_2(t) + \text{residual}
\end{equation}
and
\begin{equation}
S_n = A_n \rm{sin}(nt + \sigma_n)
\end{equation}
Here $S_1$ represents the diurnal cycle, $S_2$ the semi-diurnal cycle, $S_0$ the mean precipitation, $A_n$ the harmonic amplitude, $\sigma_n$ the phase and t is local solar time expressed in radians (i.e., $t = 2\pi t_1/24$, where $t_1$ is LST in hours).

\subsection*{CMIP6 AMIP and historical runs}
The CMIP6 data used in this study were obtained from Google's Public Dataset program stored on Google Cloud Storage.
\subsubsection*{AMIP runs}
For the analysis of monthly mean precipitation, we used the following AMIP models (all with member ID r1i1p1f1): GFDL-ESM4, GFDL-CM4, GFDL-AM4, GISS-E2-1-G, IPSL-CM6A-LR, MIROC6, BCC-CSM2-MR, BCC-ESM1, MRI-ESM2-0, CESM2, SAM0-UNICON, CESM2-WACCM, FGOALS-f3-L, CanESM5, INM-CM4-8, EC-Earth3-Veg, INM-CM5-0, MPI-ESM-1-2-HAM, NESM3, CAMS-CSM1-0, MPI-ESM1-2-HR, EC-Earth3, KACE-1-0-G, MPI-ESM1-2-LR, NorESM2-LM, E3SM-1-0, NorCPM1, FGOALS-g3, ACCESS-ESM1-5, TaiESM1, FIO-ESM-2-0, CAS-ESM2-0, CESM2-FV2, CESM2-WACCM-FV2, CMCC-CM2-SR5, EC-Earth3-AerChem, and IITM-ESM. CIESM was excluded from the analysis due to large biases.

For 3-hourly precipitation in Figs.~\ref{fig:diurnal_cycle}, S8, S16, S17, and S18, we used GFDL-CM4 (r1i1p1f1) AMIP run.

For the analysis of global mean temperature in Figs.~S19 and S29, we used the same 22 AMIP models as in\cite{kochkov2024neural}. Specifically, we used the following 17 models with the member ID r1i1p1f1: BCC-CSM2-MR, CAMS-CSM1-0, CESM2, CESM2-WACCM, CanESM5, EC-Earth3, EC-Earth3-Veg, FGOALS-f3-L, GFDL-AM4, GFDL-CM4, GFDL-ESM4, GISS-E2-1-G, IPSL-CM6A-LR, MIROC6, MRI-ESM2-0, NESM3, and SAM0-UNICON. For the remaining five models, we used alternative member IDs: r1i1p1f2 for CNRM-CM6-1 and CNRM-ESM2-1, r2i1p1f3 for HadGEM3-GC31-LL, r1i1p1f3 for HadGEM3-GC31-MM, and r1i1p1f2 for UKESM1-0-LL.

\subsubsection*{Historical runs}
Due to the limited availability of 3-hourly or daily precipitation data for AMIP models in Google's Public Dataset program, we used historical simulations for analyses requiring these temporal resolutions. In Figs.~\ref{fig:hovmoller}, \ref{fig:diurnal_cycle}, and \ref{fig:rx1day_precip_and_distribution}, we used GFDL-CM4, IPSL-CM6A-LR, BCC-CSM2-MR, MRI-ESM2-0, and GFDL-ESM4 (with member ID r1i1p1f1), as well as CNRM-CM6-1, GISS-E2-1-G, and CNRM-ESM2-1 (with member ID r1i1p1f2).

Although SST conditions are not prescribed in historical simulations, we do not expect this to qualitatively affect the results presented in these figures.

\subsection*{Comparison with observation-based data.}
To evaluate the representation of precipitation in simulations, we primarily used the Integrated Multi-satellitE Retrievals for Global Precipitation Measurement (IMERG) dataset\cite{huffman2020integrated}, which provides precipitation estimates at a $0.1^{\circ}$ spatial resolution and 30-minute temporal resolution for the period 2001–2023. This dataset utilizes data from the Global Precipitation Measurement (GPM) satellite constellation and other
data, including monthly surface precipitation gauge analyses,. To obtain a spatial resolution comparable to that of NeuralGCM, the data were conservatively regridded from the original $0.1^{\circ}$ resolution to a $2.8^\circ$ grid and averaged over time to provide 3-hourly, 6-hourly, and daily precipitation rates.

IMERG provides instantaneous estimates of precipitation (rather than cumulative values) every 30 minutes. We converted these to accumulated quantities, taking into account the IMERG documentation's suggestion: ``it is usually best to assume that this rate applies for the entire half-hour period'' (https://gpm.nasa.gov/resources/faq/how-intensity-precipitation-distributed-within-given-data-value-imerg).  However, IMERG provides these instantaneous values at some point within the 30-minute interval after the timestamp.  When time-aggregating the data, we assumed that the Y-minute accumulation rate at time X is calculated by taking the IMERG values at times [X-Y+30 min, X-Y+60 min, ..., X]. This calculation potentially shifts the accumulation by up to 30 minutes.
This shift could slightly affect the weather evaluation scores but should not significantly impact climate-related plots. To verify the robustness of our weather evaluations, we also evaluated our ensemble weather forecast against the Global Precipitation Climatology Project (GPCP) dataset (Fig.~S5) and found similar results to those obtained using IMERG.

Our analysis also incorporates the Global Precipitation Climatology Project\cite{huffman2023new} (GPCP) One-Degree Daily dataset, which provides precipitation estimates by merging data from multiple satellite sources, and surface rain gauge measurements. Over land, these satellite-based estimates are further refined using monthly rain gauge measurements. We also conservatively regrid this dataset to a $2.8^\circ$ grid.

\subsection*{Brier Scores}
We compute Brier scores comparing the (50 member) ensemble tail probabilities with observational data sets. To do this, we first compute thresholds $t_i$, corresponding to quantiles 
$q_i=(0.95, 0.99)$
(separately for every latitude/longitude/dayofyear). In other words, with $Y$ ground truth, the historical $\mathrm{P}[Y < t_i] = q_i$. The Brier score at each latitude/longitude/lead-time is then defined with an average over initial times $\mathcal{T}$ as
\begin{align*}
    \frac{1}{|\mathcal{T}|}\sum_{t\in\mathcal{T}}
    \left|
    \frac{1}{50}\sum_{n=1}^{50} \mathbf{1}_{X^{(n)}_t > t_i}
    - \mathbf{1}_{Y_t > t_i}
    \right|^2
\end{align*}
Above, $\mathbf{1}_{X_t > t_i}=1$ when $X_t > t_i$ and $=0$ when $X_t \leq t_i$, $\{X_t^{(1)},\ldots X_t^{(50)}\}$ is the $50-$member ensemble forecast value at the latitude/longitude/lead-time, (implying initial + lead time is $t$), and $Y_t$ is the corresponding ground truth. 

\subsection*{Probabilistic climatological forecasts}
As an additional baseline, we generate a size 50 ensemble of forecasts $X_{clim}$ by sampling historical IMERG data $X_{hist}$. Creation of the forecast at initial time $t$ starts by choosing a random source initial time $s$. The forecast at lead time $\tau$ is then $X_{clim}(t+\tau) = X_{hist}(s + \tau)$. To choose the initial time $s$, we first choose \texttt{s.year} uniformly in $1990 - 2019$ (for ERA5) and $2001-2019$ (for IMERG). Second, we choose \texttt{s.dayofyear} uniformly in \texttt{[t.dayofyear - 7, t.dayofyear + 7]}. Time of day is unchanged and sampling is done without replacement.

\bibliographystyle{sn-nature.bst}
\bibliography{precip-bibliography}
\end{document}


\title[NeuralGCMs]{Neural general circulation models optimized to predict satellite-based precipitation observations - Supplementary Information}


\author*[1]{\fnm{Janni} \sur{Yuval}}
\email{janniyuval@google.com}
\equalcont{These authors contributed equally to this work.}
\author[1]{\fnm{Ian} \sur{Langmore}}
\equalcont{These authors contributed equally to this work.}
\author[1]{\fnm{Dmitrii} \sur{Kochkov}}
\equalcont{These authors contributed equally to this work.}
\author[1]{\fnm{Stephan} \sur{Hoyer}}
\equalcont{These authors contributed equally to this work.}
\affil[1]{\orgname{Google Research, Mountain View, CA}}

\maketitle


\subsection*{Limitations of current NeuralGCM model}
There several notable caveats in the current version of NeuralGCM which we highlight here: 

\textbf{Temperature bias and spread of simulations}. 
The current version of NeuralGCM exhibits a global mean temperature bias of approximately 0.5K at 850hPa  (Fig. \ref{sifig:global_mean_tempearture}). While significant, this bias is smaller than that in some AMIP models. Another key issue with NeuralGCM precipitation model, also present to some extent in the previous version, is the large spread in global mean temperature within the NeuralGCM ensemble. This spread is substantially larger than that obtained in physics-based models, such as MIROC6 (Fig.~\ref{sifig:global_mean_tempearture}). Interestingly, we found in Neural-evap, which was trained to predict evaporation and diagnose precipitation, this spread is substantially smaller (Fig.~\ref{sifig:global_mean_tempearture_evap}).

\textbf{Unrealistic instantaneous evaporation}
The NeuralGCM presented here was trained to optimize both 6-hourly evaporation rates and accumulated precipitation.  For the model described in the main text, where precipitation is predicted and evaporation diagnosed, the annual mean diurnal cycle of evaporation is consistent with that of ERA5 (Fig.~\ref{sifig:mean_evaporaiton}). However, unrealistic artifacts are evident in instantaneous snapshots of evaporation (Fig.~\ref{sifig:instantenous_evaporaiton}). This likely arises from the indirect estimation of evaporation, which hinders the model's ability to achieve smooth evaporation fields. Notably, these artifacts are absent in the NeuralGCM-evap (configuration where evaporation is predicted and precipitation diagnosed; Fig.~\ref{sifig:instantenous_evaporaiton}).
As discussed earlier, NeuralGCM is trained on potentially conflicting datasets. This may further contribute to the challenges in obtaining consistent values for both evaporation and precipitation simultaneously.

\textbf{Unrealistic precipitation at frequencies higher than 6 hours}
As was discussed in the manuscript and was shown in Figs.~6, \ref{sifig:global_mean_precipitation} , the NeuralGCM model that is introduced in the manuscript exhibits unrealistic diurnal features, with  certain  times  of  day  experiencing significantly  more  precipitation  than  others. Therefore, we recommend against using this configuration at frequencies higher than 6-hourly. These issues, however, do not occur in NeuralGCM-evap (Figs.~\ref{sifig:global_mean_precipitation},\ref{sifig:diurnal_cycle_evap}).

\textbf{Stability overall dependent on the initial seed.}
We introduced a NeuralGCM (and NeuralGCM-evap) model capable of stable 20-year simulations for all 37 initial conditions tested (and 731/732 simulations ran 22 years without instability). However, during development, we observed that models trained with different random seeds (resulting in different initial model parameters or weights) exhibited significant variations in stability. Specifically, models trained with similar settings but initialized with different seeds demonstrated markedly different stability durations, despite their similar training. We note that improving stability of hybrid models remains a challenging task.

\subsection*{Key changes in NeuralGCM to improve stability}
We observed that some instabilities in the previous version of NeuralGCM were related to a drift in the global mean log surface pressure. We found that a similar drift also occurs when running NeuralGCM dynamical core alone with realistic orography (and without the ML parameterization). Furthermore, we found that NeuralGCM models exhibit some sensitivity to initial conditions, even in long integrations. For example, models initialized at different dates could produce slightly different mean temperatures. To mitigate these issues, we allowed the neural network parameterization to modify the log surface pressure prognostic variable and constrained the global mean log surface pressure to remain constant during long integrations. Additionally, we removed the stochastic component of the encoder (excluded the random field input), which further improved stability.

\subsubsection*{Other changes in NeuralGCM precipitation model}
Besides the configuration changes mentioned above, there are two minor differences between the NeuralGCM precipitation model presented here and the one described in\cite{kochkov2024neural}:

\begin{itemize}
    \item We use a dynamical core time step of 20 minutes (compared to 12 minutes in the $2.8^\circ$ model presented in \cite{kochkov2024neural}).
    \item We include cloud fields as inputs to the decoder, which were mistakenly omitted in the previous version.
\end{itemize}

\subsection*{Dispersion relation}
Fig.~\ref{sifig:Wheeler_Kiladis} presents the dispersion relationships in the zonal wavenumber-frequency domain, computed following the methodology of Wheeler and Kiladis\cite{wheeler1999convectively} using the wavenumber-frequency Python package\cite{Madieros_wn}. 
Both the symmetric and anti-symmetric components of the Wheeler-Kiladis diagram for NeuralGCM show a close resemblance to the observed dispersion relations, capturing key tropical wave modes such as the Madden-Julian Oscillation (MJO), Kelvin waves, and equatorial Rossby waves. This representation aligns more closely with IMERG data than the GFDL model, although NeuralGCM exhibits an unrealistic peak at low frequencies. Notably, the dispersion relation derived from ERA5 demonstrates an even closer correspondence to IMERG than NeuralGCM.

\subsection*{Results for NeuralGCM evaporation model}
We also experimented with a model that uses a neural network to predict evaporation:
\begin{equation}\label{eq:e_with_nn}
    E = {\rm{NN}_{\rm{evap}}}(X),
\end{equation}
where $\text{NN}_\text{evap}$ is a neural network that predicts evaporation ($E$), and $X$ represents the inputs to the network.
Precipitation ($P$) is then diagnosed by enforcing water conservation in the column (Eq.~1 in Methods):
\begin{equation}\label{eq:p_from_constraint}
    P =  \frac{1}{g}\int_{0}^{1} \sum_{\rm{i}} \left(\frac{dq}{dt}\right)_{\rm{i}}^{\rm{NN_{\rm{tend}}}} p_{\rm{s}} d\sigma + {\rm{NN_{\rm{evap}}}}(X).
\end{equation}
We refer to this model as NeuralGCM-evap.
This formulation has the advantage of being more aligned with how current atmospheric models are constructed, where evaporation is calculated using a surface scheme. Furthermore, the neural network employed in this approach is smaller, as it only takes near-surface atmospheric values as input (see below).

One significant disadvantage of this model is that it can produce negative precipitation values. While we attempted to address this issue, techniques that ensured non-negative precipitation (while predicting evaporation) led to less stable models. 

\subsubsection*{Results for evaporation model}
We evaluated NeuralGCM-evap following the same procedure described in the main text, conducting 20-year simulations with 37 different initial conditions. 36 out of 37 initial conditions remained stable and exhibited no drift in global mean temperature (Fig.~\ref{sifig:global_mean_tempearture_evap}).

Overall, the results from this model are also quite compelling, with the major caveat that it produces negative precipitation values (see the precipitation rate frequency distribution in Fig.~\ref{sifig:distributions_precip_rate_evap}). In terms of mean precipitation (Fig.~\ref{sifig:mean_precip_evap}), extreme precipitation (Fig.~\ref{sifig:extreme_precip_evap}) and diurnal cycle (Fig.~\ref{sifig:diurnal_cycle_evap}) this model exhibits even better performance than the model presented in the main text. However, this assessment includes negative precipitation values in the calculation of MAE. When these negative values are set to zero, the MAE for mean precipitation increases (Fig.~\ref{sifig:mean_precip_evap}c).
NeuralGCM-evap also exhibits a realistic precipitation also at frequencies higher than 6 hours which it didn't directly train on (Figs.~\ref{sifig:global_mean_precipitation}, \ref{sifig:diurnal_cycle_evap}) and does not exhibit the unrealistic features present in NeuralGCM at these higher frequencies. Furthermore, NeuralGCM-evap produces a time-space spectrum similar to that of IMERG (Fig.~\ref{sifig:Wheeler_Kiladis_evap}).

\subsubsection*{Input features for evaporation neural network}
The primary input features to the neural network that predicts evaporation include the surface values of zonal and meridional wind, temperature anomalies, and specific humidity.

Additionally, we incorporate an 8-dimensional location-specific embedding vector for each horizontal grid point. This vector, initialized with random values, is trained to represent unique geographical features. We also utilize a surface embedding network that receives surface-related inputs, specifically sea surface temperature (SST) and sea ice concentration. Over land and ice where SST data are unavailable, the lowest model level temperature and specific humidity are included as input. (Full details are provided in \cite{kochkov2024neural} which uses a similar surface embedding network.)

\subsubsection*{Evaporation network architecture}
The evaporation network, like the precipitation network, employs an Encode-Process-Decode (EPD) architecture but with a smaller size than the precipitation NN. The ``Encode'' layer maps the input features to a latent vector of size 8 (compared to 64 for the precipitation network). Each ``Process'' block utilizes a 3-layer MLP with 8 hidden units (compared to 64 for the precipitation network). Finally, the ``Decode'' layer maps the latent vector of size 8 (64 for the precipitation network) to the hourly evaporation rate.

\subsection*{Potential benefits to physical consistency}
To illustrate the potential benefits of enforcing consistency with the water budget, we compare the precipitable water (PW) distribution of our model to that of a different NeuralGCM model trained without this constraint. As baselines, we use ERA5, which our models used as a target, and AQUA AIRS measurements\cite{Barnet_airs}, which were not used in training.
The unconstrained approach, akin to typical machine learning models, adds a new output (precipitation) without direct interaction with other variables, except through the optimization process. Namely,
\begin{equation}\label{eq:p_with_nn}
    P = {\rm{NN}_{\rm{precip}}}(X)
\end{equation}
where $\text{NN}_\text{precip}$ is a neural network that predicts precipitation ($P$), and $X$ represents the inputs to the network (Methods). 
As expected, the unconstrained model produces a PW distribution similar to that of ERA5, which was used as the training target (Fig. \ref{sifig:pw_airs_constraint}).

In contrast, for the model presented in the main manuscript, we also diagnose evaporation ($E$) by enforcing water conservation in the column (Eq.~1 in Methods):
\begin{equation}\label{eq:e_from_constraint}
    E = {\rm{NN_{\rm{precip}}}}(X) - \frac{1}{g}\int_{0}^{1} \sum_{\rm{i}} \left(\frac{dq}{dt}\right)_{\rm{i}}^{\rm{NN_{\rm{tend}}}} p_{\rm{s}} d\sigma.
\end{equation}
Enforcing the water budget constraint results in feedback into the neural network for tendency prediction. This shifts the PW distribution in the NeuralGCM model towards higher values, which are observed in AQUA AIRS but are not observed in ERA5, albeit with some overestimation at the high end (Fig. \ref{sifig:pw_airs_constraint}). This comparison highlights the potential of incorporating physical constraints into machine learning models, as such constraints can influence model dynamics and lead to improved realism.

\subsection*{Extreme precipitation sensitivity to changes in global mean temperature}
Detecting statistically significant changes in extreme precipitation over relatively short periods necessitates numerous ensemble members, making computational efficiency crucial. The NeuralGCM model described here can simulate approximately 1200 years in 24 hours on a single TPU v4. This enables the feasible execution of O(1000) 20-year realizations using 16 TPU v4 days. (In practice, we use several TPUs in parallel and do not need to wait 16 days for the results.)

As a proof of concept for how NeuralGCM can be used to estimate quantities of interest, we ran 732 ensemble members for 22 years (two ensemble members initialized every 24 hours during 2001; 731 out of 732 remained stable for the full 22 years of simulations). We use this ensemble to estimate the sensitivity of annual maximum precipitation (Rx1day) to changes in global mean temperature.

We note that our analysis does not isolate the influence of anthropogenic climate change on extreme precipitation events, as we prescribe SSTs that have also been influenced by natural variability and contribute to the obtained signal. Future work will need to incorporate an ocean model to facilitate a more controlled analysis, enabling comparisons of different years with varying sea surface temperature (SST) realizations.
Previous studies have used both models \cite{kharin2013changes, pfahl2017understanding} and observations \cite{asadieh2015global, o2015precipitation} to estimate the sensitivity of extreme precipitation to changes in global mean temperature. Although there are still large uncertainties regarding the response of extreme precipitation to global mean temperature increase on regional scales, there is ample evidence from both observations and models that, on large spatial scales, there is an overall increase in extreme precipitation globally.

For each grid box, the Rx1day values from all years and ensemble members are regressed against the 850 hPa temperature anomalies (calculated separately for each year) using the Theil-Sen estimator. The regression coefficient is then divided by the mean of the annual-maximum daily precipitation rate (averaged over the whole ensemble) at that grid box to yield a sensitivity expressed in units of \% K$^{-1}$ (Fig. \ref{sifig:sensitivity_Rx1day}). This methodology is similar to that used previously \cite{o2015precipitation}, with the difference that we use the 850 hPa temperature instead of the near-surface temperature (since NeuralGCM does not output near surface temperature).

Overall, the results shown in Fig.~\ref{sifig:sensitivity_Rx1day} are mostly consistent with those obtained from the ensemble mean of CMIP5 models \cite{kharin2013changes,pfahl2017understanding}, where there is an overall increase in extreme precipitation in most regions (particularly near the equator in the Pacific). However, there are several regions exhibiting a decrease in Rx1day with increasing global mean temperature, including subtropical regions in the North Atlantic and South Pacific near South America, as well as North Africa and western Australia. There are also some clear differences; for example, NeuralGCM indicates a decrease in the Arabian Peninsula and the Arabian Sea, which is not seen in the ensemble mean of CMIP5 models.

We emphasize that the differences between NeuralGCM and the ensemble mean of CMIP5 models are not necessarily a negative indication, since the true sensitivity is unknown and because an ensemble mean of models attenuates extreme values that occur in individual models.

To make a rough comparison between NeuralGCM and observations over land, we calculate the median sensitivity for each latitudinal band. Similar to observations, we find that over land, all latitudes show an increase in Rx1day with increasing temperature (Fig.~\ref{sifig:sensitivity_Rx1day}c). Confidence intervals are estimated using 1000 smoothed bootstrap\cite{Young1990-ym} iterations at each latitude. First, locations are sampled with replacement. Second, for each location, we randomly draw a slope value from a normal distribution centered at the location's mean slope, and standard deviation obtained from the Theil-Sen regression. The normal sampling step is a smoothing\cite{Young1990-ym}, which improves the estimate in light of the limited number of points at each latitude. The global mean sensitivity is 4.2\% K$^{-1}$ (with 95\% confidence intervals calculated from the Theil-Sen regression of 3.9–4.5\% K$^{-1}$), which is slightly lower than the range obtained from some previous work, estimated to be 5–10\% K$^{-1}$ \cite{westra2013global, o2015precipitation} (with some sensitivity to the exact method of estimation), but still within a reasonable range. We note that observations have a very non-uniform distribution over land, so the comparison is not exactly apples-to-apples. Furthermore, previous work indicates that extreme precipitation responses to global mean temperature are more modest at coarser resolutions \cite{lu2014robust} and lower percentiles \cite{tabari2020climate}. It is possible that at higher resolutions or higher percentiles, the response would be more pronounced. To facilitate a reasonable comparison to observations, we mask out regions with Rx1day $<$ 20 mm/day, which effectively masks out North Africa, where the observational record is practically non-existent.

\renewcommand{\thefigure}{S\arabic{figure}} 


\begin{figure*}
\begin{center}
\makebox[\textwidth]{\colorbox{white}{\includegraphics[width=0.65\paperwidth]{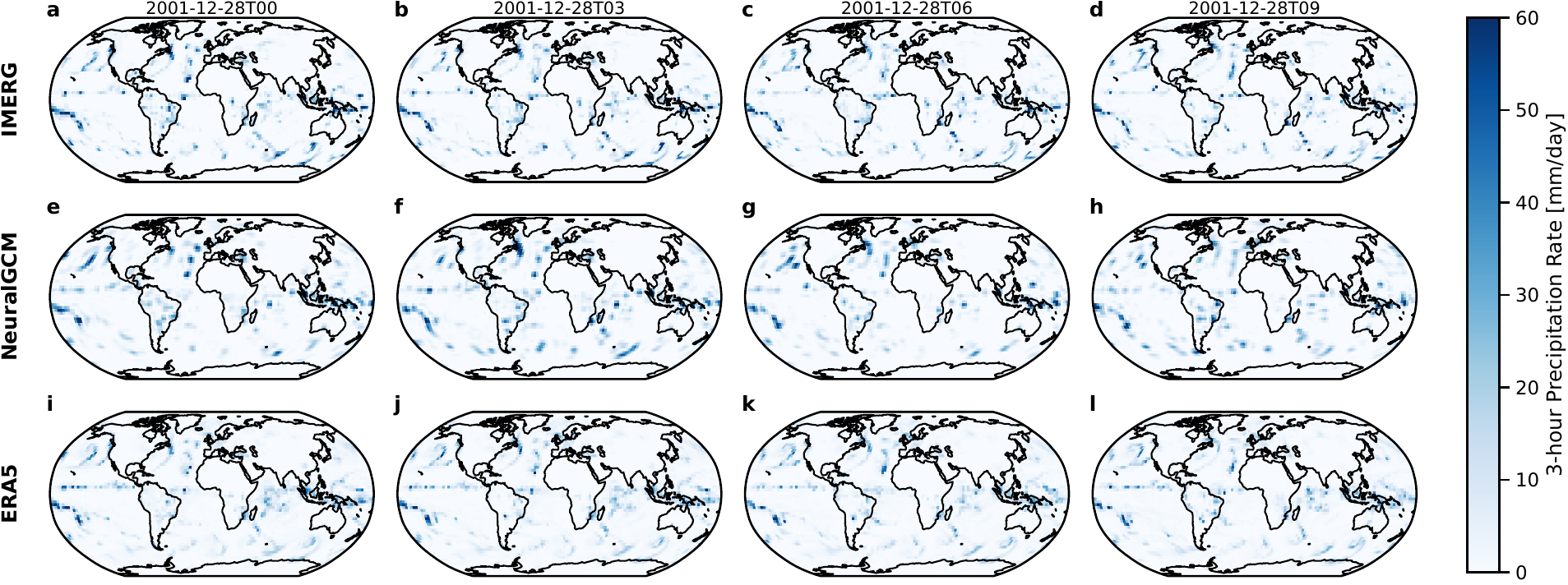}}}
\end{center}
\caption{
3-Hourly precipitation rate from IMERG, NeuralGCM and ERA5.
NeuralGCM was initialized on 12-27-2001 (24 hours before the first snapshot is shown).
}\label{sifig:forecast_3_hour_snpashots}
\end{figure*}

\begin{figure*}
\begin{center}
\makebox[\textwidth]{\colorbox{white}{\includegraphics[width=0.5\paperwidth]{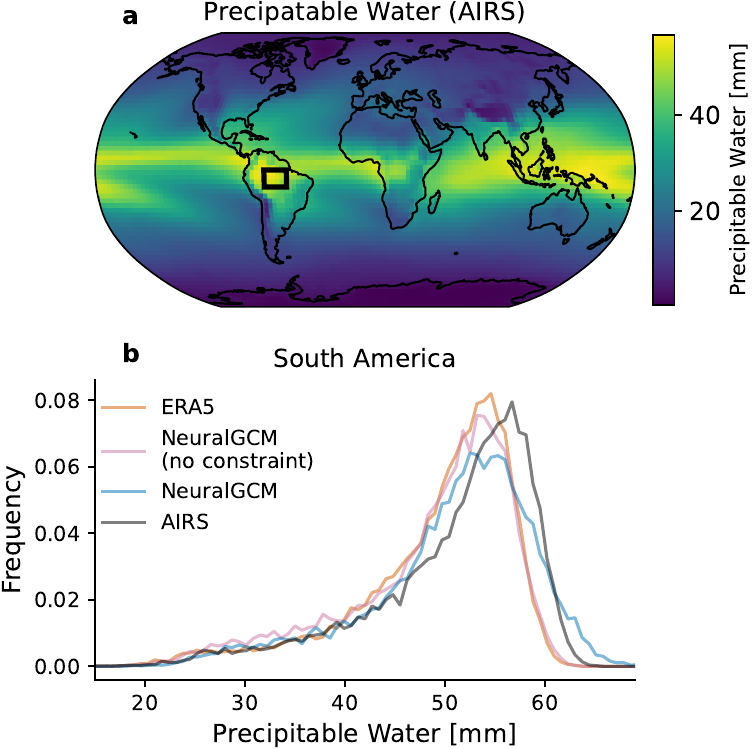}}}
\end{center}
\caption{
Comparison of precipitable water (PW) from different datasets and models. (a) Time-averaged PW (excluding cloud water) from AQUA AIRS\cite{Barnet_airs} measurements at 1:30 PM local time, averaged over 2018–2019. (b) Frequency distribution of PW over South America (latitudes $-10^\circ$ to $0\circ$, longitudes $292.5^\circ$ to $307.5^\circ$) for ERA5, AQUA AIRS, and two NeuralGCM configurations: (1) ``no constraint'' – trained to predict precipitation without any constraint on evaporation or the water budget; (2) NeuralGCM model discussed in the manuscript which uses water budget constraint, with optimization of both precipitation and evaporation.  AIRS data are from 1:30 PM local time.  For ERA5 and NeuralGCM, data are from 19:00 UTC, which roughly corresponds to 1:30 PM local time in the chosen region. NeuralGCM simulations were initialized on January 1, 2018, and run for two years.
}\label{sifig:pw_airs_constraint}
\end{figure*}

\begin{figure*}
\begin{center}
\makebox[\textwidth]{\colorbox{white}{\includegraphics[width=0.65\paperwidth]{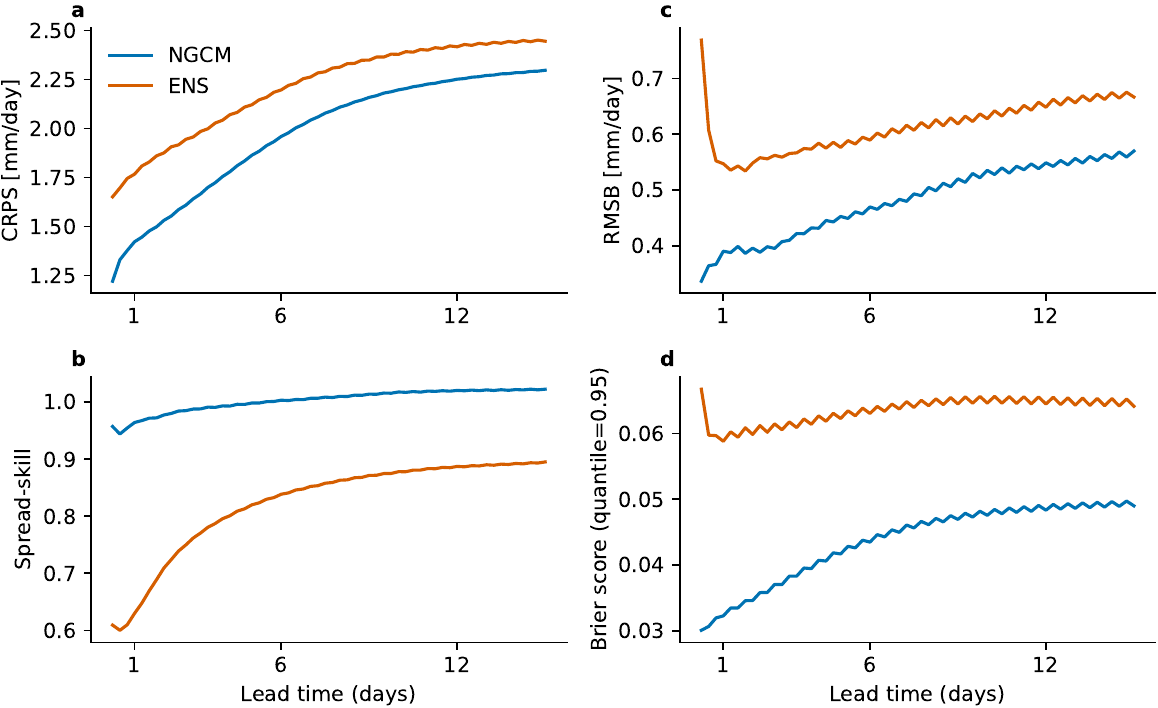}}}
\end{center}
\caption{
Precipitation forecasting accuracy scores for 6-hour accumulated precipitation, evaluated against IMERG. Area-weighted mean, calculated over all longitudes and latitudes between $-60^\circ$ to $60^\circ$ for: (a) Continuous Ranked Probability Score (CRPS). (b) Spread-skill ratio. (c) Root-mean-square bias (RMSB). (d) Brier score (0.95 quantile). Comparisons are shown for NeuralGCM and ECMWF ensemble. 
}\label{sifig:evals_6h}
\end{figure*}

\begin{figure*}
\begin{center}
\makebox[\textwidth]{\colorbox{white}{\includegraphics[width=0.65\paperwidth]{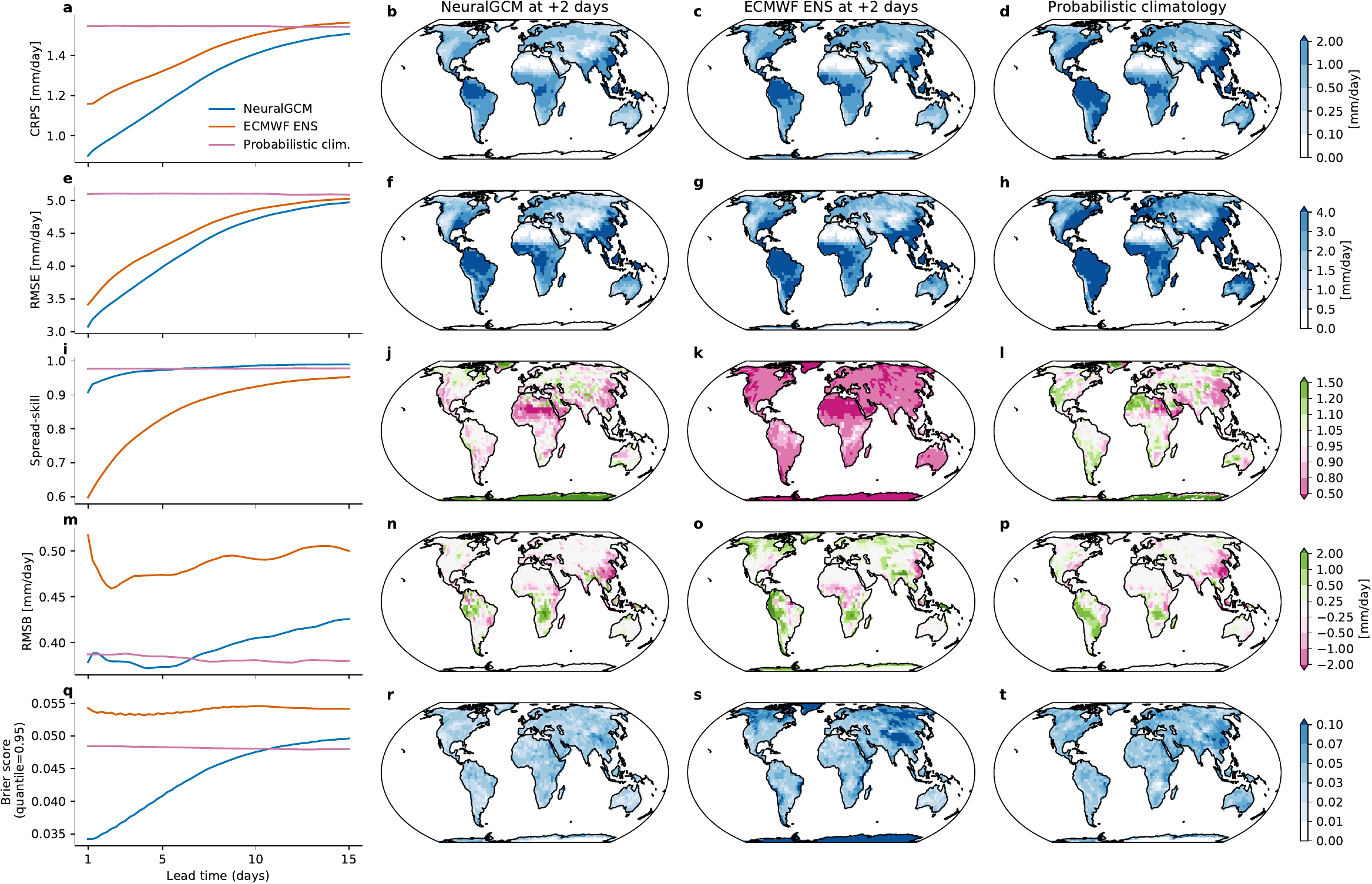}}}
\end{center}
\caption{
Precipitation forecasting accuracy scores for 24-hour accumulated precipitation over land, evaluated against IMERG. This figure is similar to Fig.~2, but the evaluation here is restricted to land areas.
}\label{sifig:land_evals_24h_spatial}
\end{figure*}

\begin{figure*}
\begin{center}
\makebox[\textwidth]{\colorbox{white}{\includegraphics[width=0.65\paperwidth]{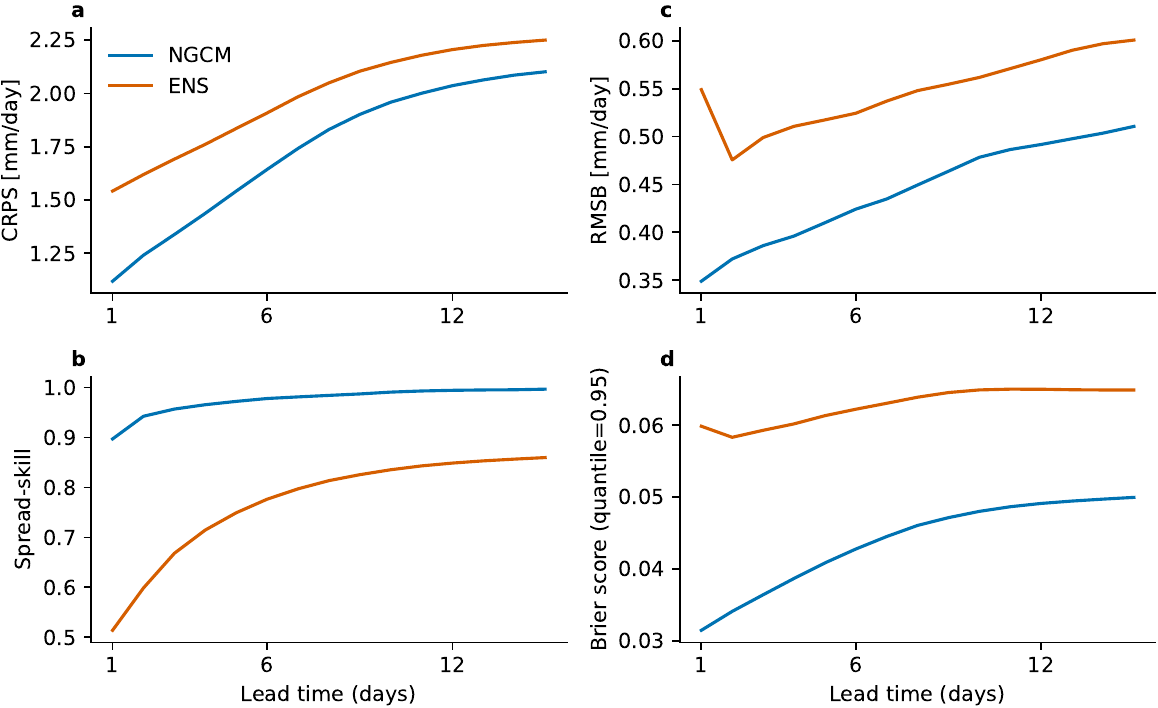}}}
\end{center}
\caption{
Precipitation forecasting accuracy scores for 24-hour accumulated precipitation, evaluated against GPCP.
like Fig.~\ref{sifig:evals_6h}, but using GPCP as ground truth.
}\label{sifig:weather_evals_gpcp}
\end{figure*}

\begin{figure*}
\begin{center}
\makebox[\textwidth]{\colorbox{white}{\includegraphics[width=0.65\paperwidth]{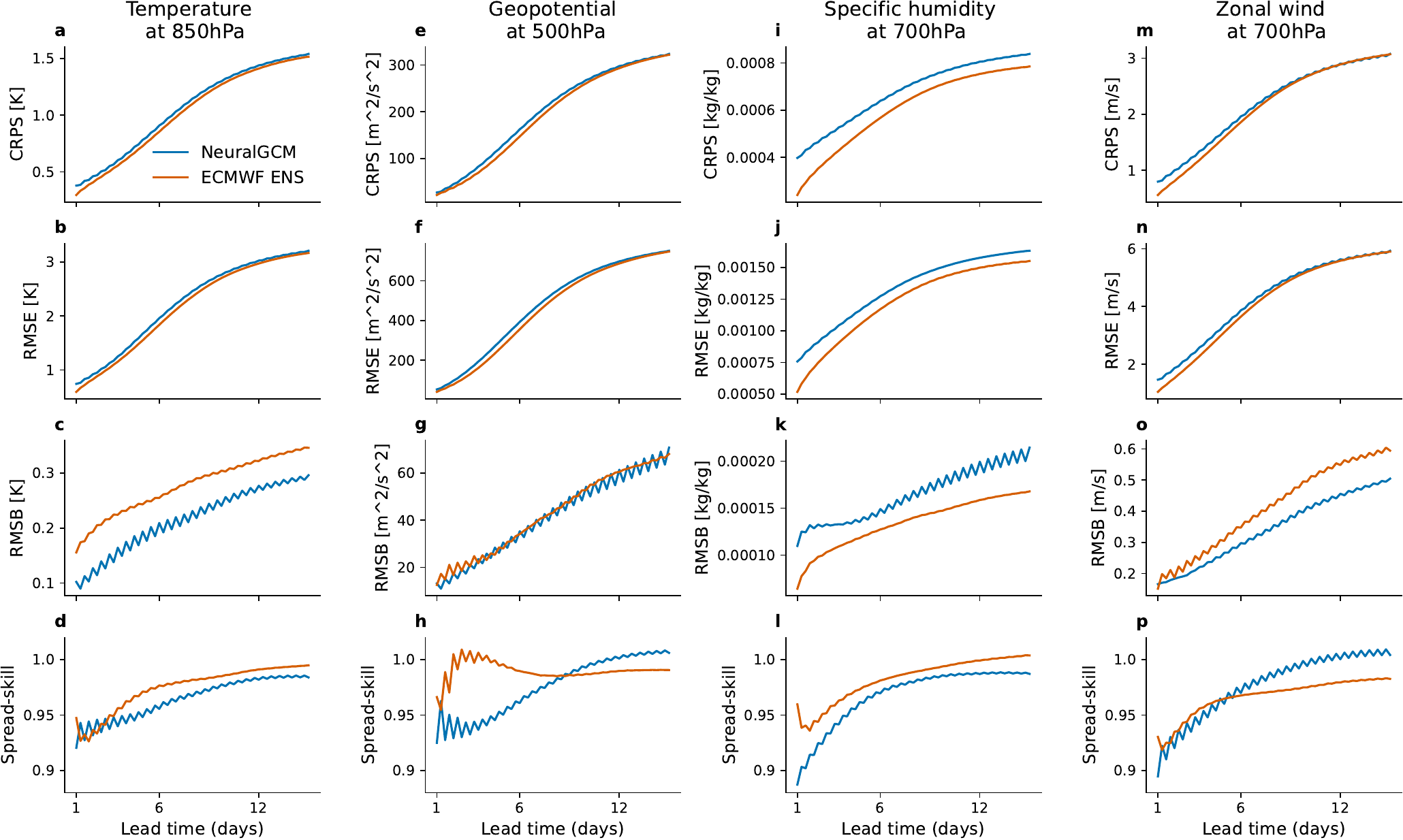}}}
\end{center}
\caption{
Weather forecasting accuracy scores for NeuralGCM and the ECMWF ENS for various atmospheric variables. Rows show different skill metrics: (1) Continuous Ranked Probability Score (CRPS), (2) ensemble mean root-mean-square error (RMSE), (3) root-mean-square bias (RMSB), and (4) spread-skill ratio. Columns show different variables: (a-d) temperature at 850 hPa, (e-h) geopotential at 500 hPa, (i-l) specific humidity at 700 hPa, and (m-p) zonal wind at 700 hPa. NeuralGCM is compared with ERA5 as the ground truth, whereas ECMWF-ENS is compared with the ECMWF operational analysis (that is, HRES at 0-hour lead time), to avoid penalizing the operational forecasts for different biases than ERA5.
}\label{sifig:weather_eval_weather_vars}
\end{figure*}

\begin{figure*}
\begin{center}
\makebox[\textwidth]{\colorbox{white}{\includegraphics[width=0.45\paperwidth]{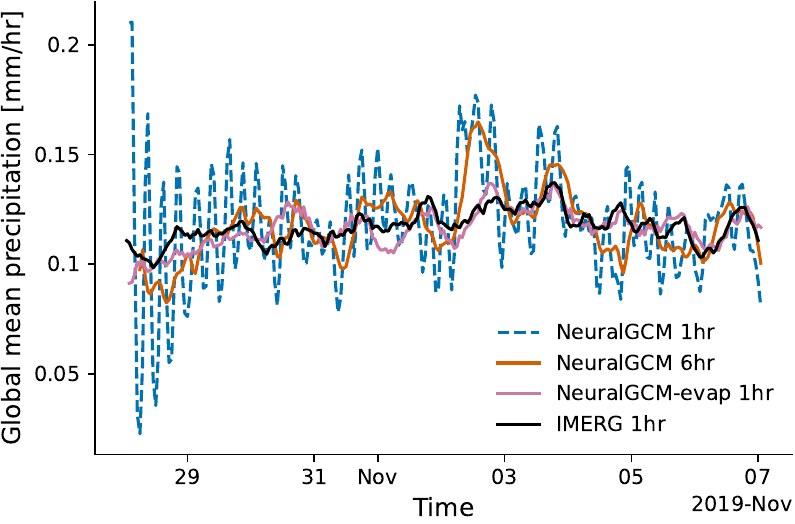}}}
\end{center}
\caption{
Global mean precipitation for 10-day forecasts. Time series of global mean precipitation for NeuralGCM at 1-hour and 6-hour frequencies, IMERG at 1-hour frequency, and NeuralGCM-evap at 1-hour frequency. NeuralGCM exhibits unrealistic fluctuations at frequencies higher than 6-hourly (see also Fig. 6 in the manuscript), while NeuralGCM-evap does not (see also Fig. \ref{sifig:diurnal_cycle_evap}). NeuralGCM forecasts were initialized on October 28, 2019.
}\label{sifig:global_mean_precipitation}
\end{figure*}

\begin{figure*}
\begin{center}
\makebox[\textwidth]{\colorbox{white}{\includegraphics[width=0.65\paperwidth]{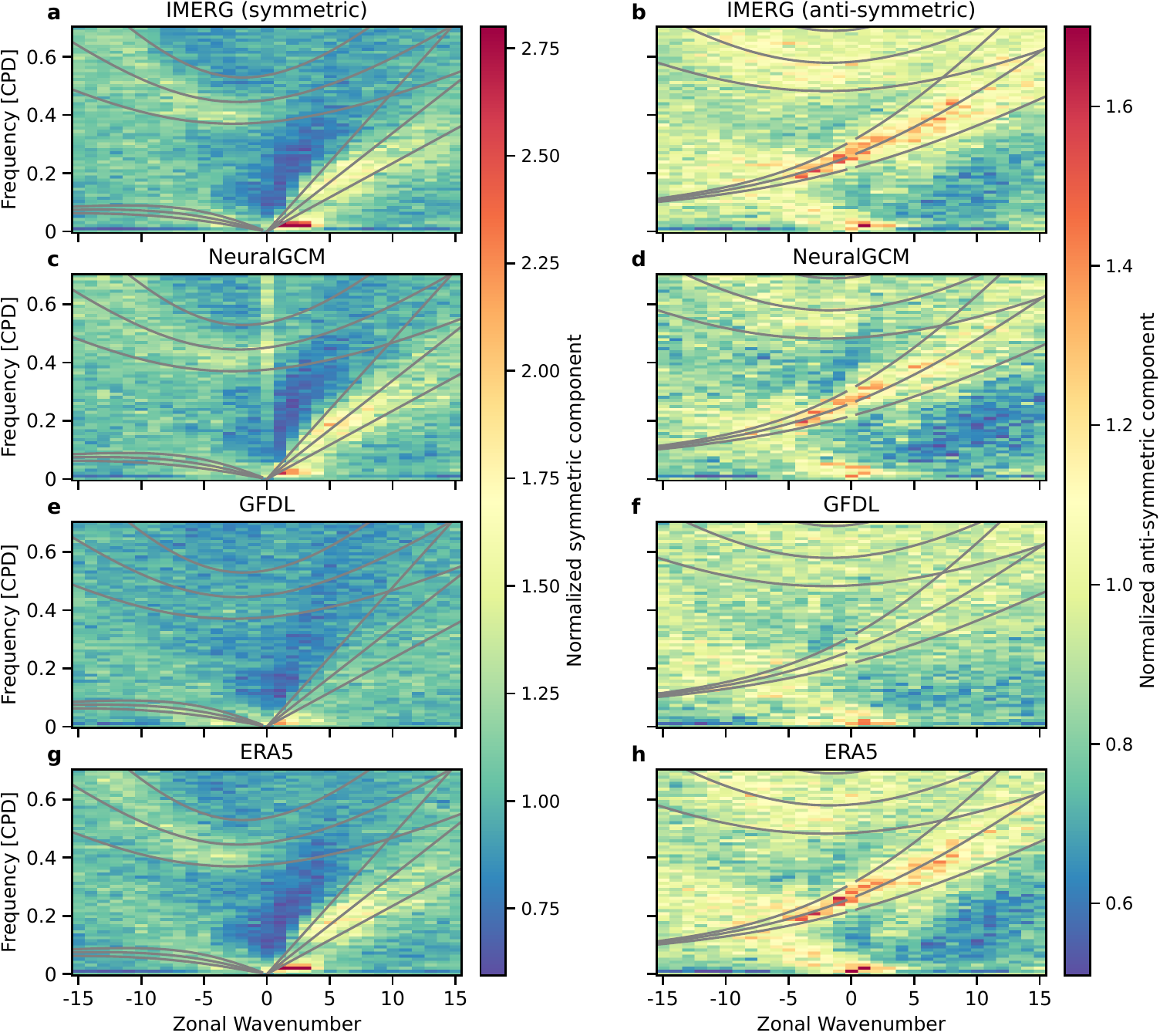}}}
\end{center}
\caption{
Space-time spectra of precipitation for IMERG, NeuralGCM, GFDL (AMIP run), and ERA5 (2002–2014).  The Wheeler-Kiladis diagrams were constructed using 96-day windows with a 60-day overlap. To highlight the dominant wave modes, the power spectrum was normalized by a smoothed background spectrum, which was estimated by repeatedly convolving the spectrum in the frequency dimension (separately for positive and negative frequencies).
}\label{sifig:Wheeler_Kiladis}
\end{figure*}


\begin{figure*}
\begin{center}
\makebox[\textwidth]{\colorbox{white}{\includegraphics[width=0.65\paperwidth]{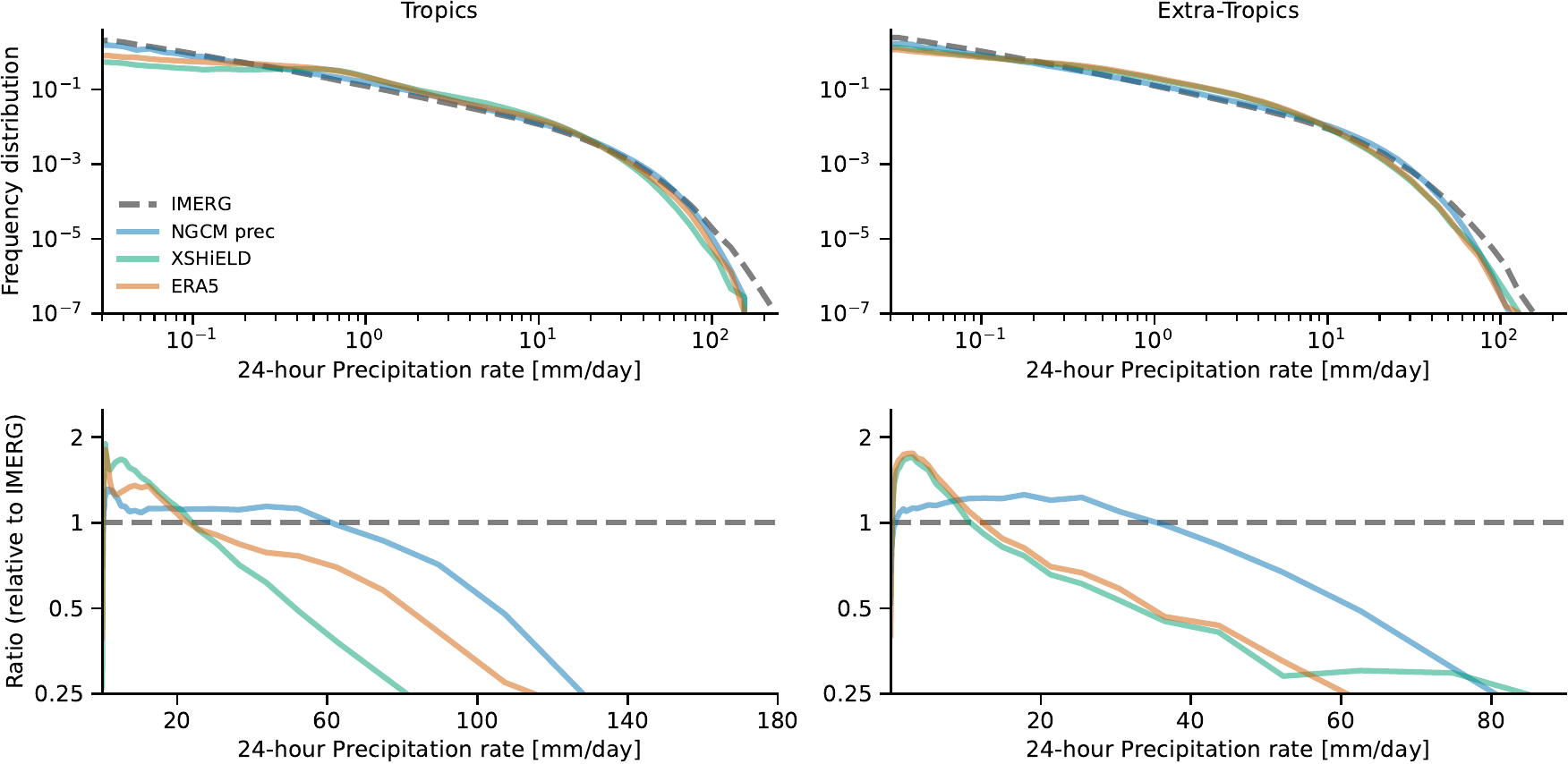}}}
\end{center}
\caption{
Precipitation rate distributions for IMERG, NeuralGCM, ERA5 and X-SHEiLD.
Frequency distribution of 24-hourly precipitation rate for (a) tropics (latitudes -20 to 20) and (b) extra-tropics (latitudes 30 to 70 in both hemispheres) and the relative distribution (normalized by the IMERG value) for (c) tropics and (d) extra-tropics. 
Distributions for all models are calculated from the dates available in X-SHEiLD run (January 18th 2020 to January 17th 2021). NeuralGCM model was initialized on 2001-12-27. All models coarsened to $2.8^{\circ}$ resolution.
}\label{sifig:distributions_precip_rate}
\end{figure*}

\begin{figure*}
\begin{center}
\makebox[\textwidth]{\colorbox{white}{\includegraphics[width=0.65\paperwidth]{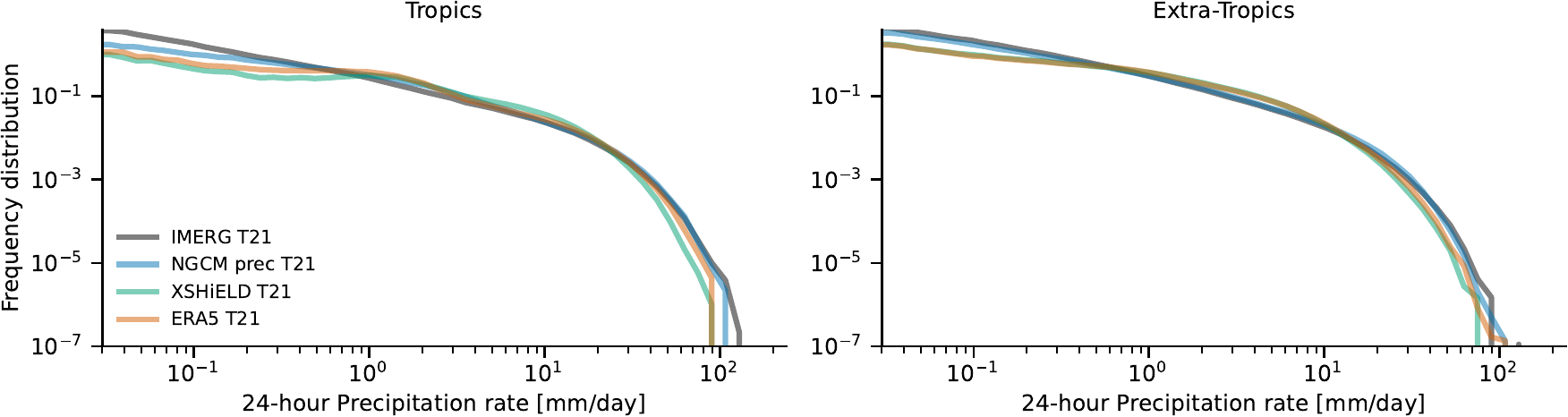}}}
\end{center}
\caption{
Precipitation rate distributions for IMERG, NeuralGCM, ERA5 and X-SHEiLD coarsened to $5.6^{\circ}$ resolution. 
This figure is similar to Fig.~\ref{sifig:distributions_precip_rate} (a-b), but all models were coarse grained to $5.6^{\circ}$ resolution.
}\label{sifig:distributions_precip_rate_t21}
\end{figure*}


\begin{figure*}
\begin{center}
\makebox[\textwidth]{\colorbox{white}{\includegraphics[width=0.65\paperwidth]{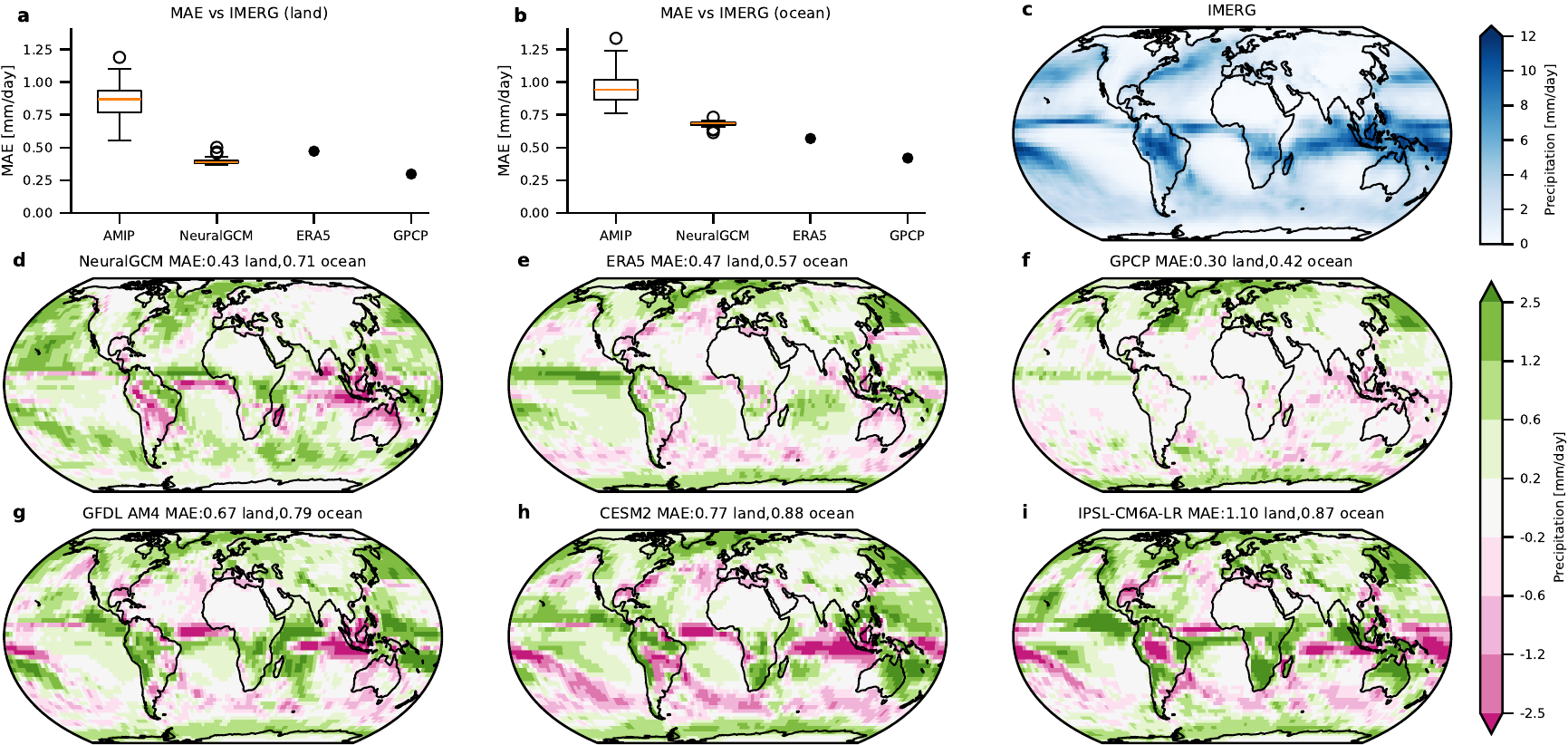}}}
\end{center}
\caption{
Mean bias in precipitation averaged over 2002–2014 for December-January-February (same as Fig.~4 but for DJF) 
}\label{sifig:mean_precip_DJF}
\end{figure*}

\begin{figure*}
\begin{center}
\makebox[\textwidth]{\colorbox{white}{\includegraphics[width=0.65\paperwidth]{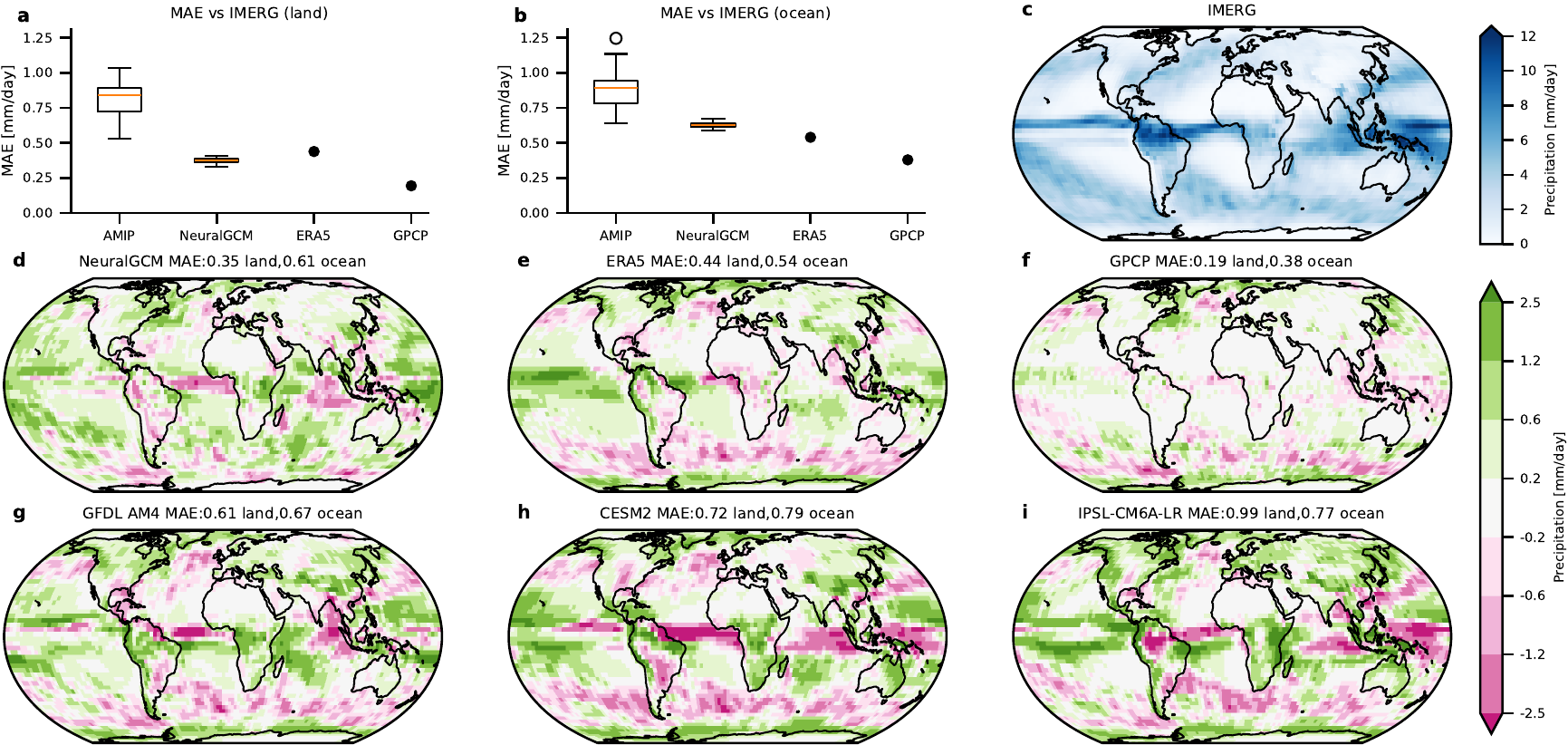}}}
\end{center}
\caption{
Mean bias in  precipitation averaged over 2002–2014 for March-April-May (same as Fig.~4 but for MAM) 
}\label{sifig:mean_precip_MAM}
\end{figure*}

\begin{figure*}
\begin{center}
\makebox[\textwidth]{\colorbox{white}{\includegraphics[width=0.65\paperwidth]{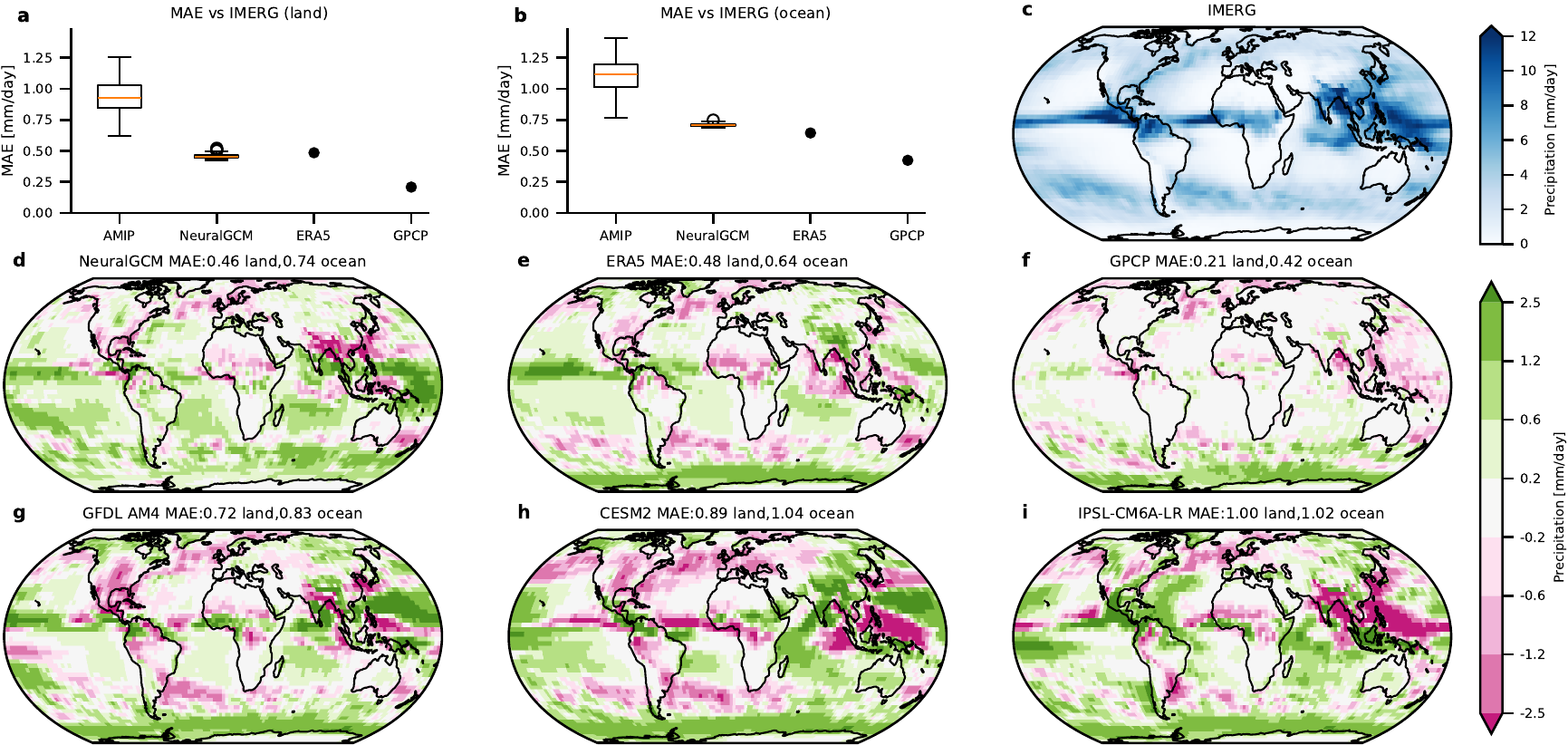}}}
\end{center}
\caption{
Mean bias in  precipitation averaged over 2002–2014 for June-July-August (same as Fig.~4 but for JJA) 
}\label{sifig:mean_precip_JJA}
\end{figure*}

\begin{figure*}
\begin{center}
\makebox[\textwidth]{\colorbox{white}{\includegraphics[width=0.65\paperwidth]{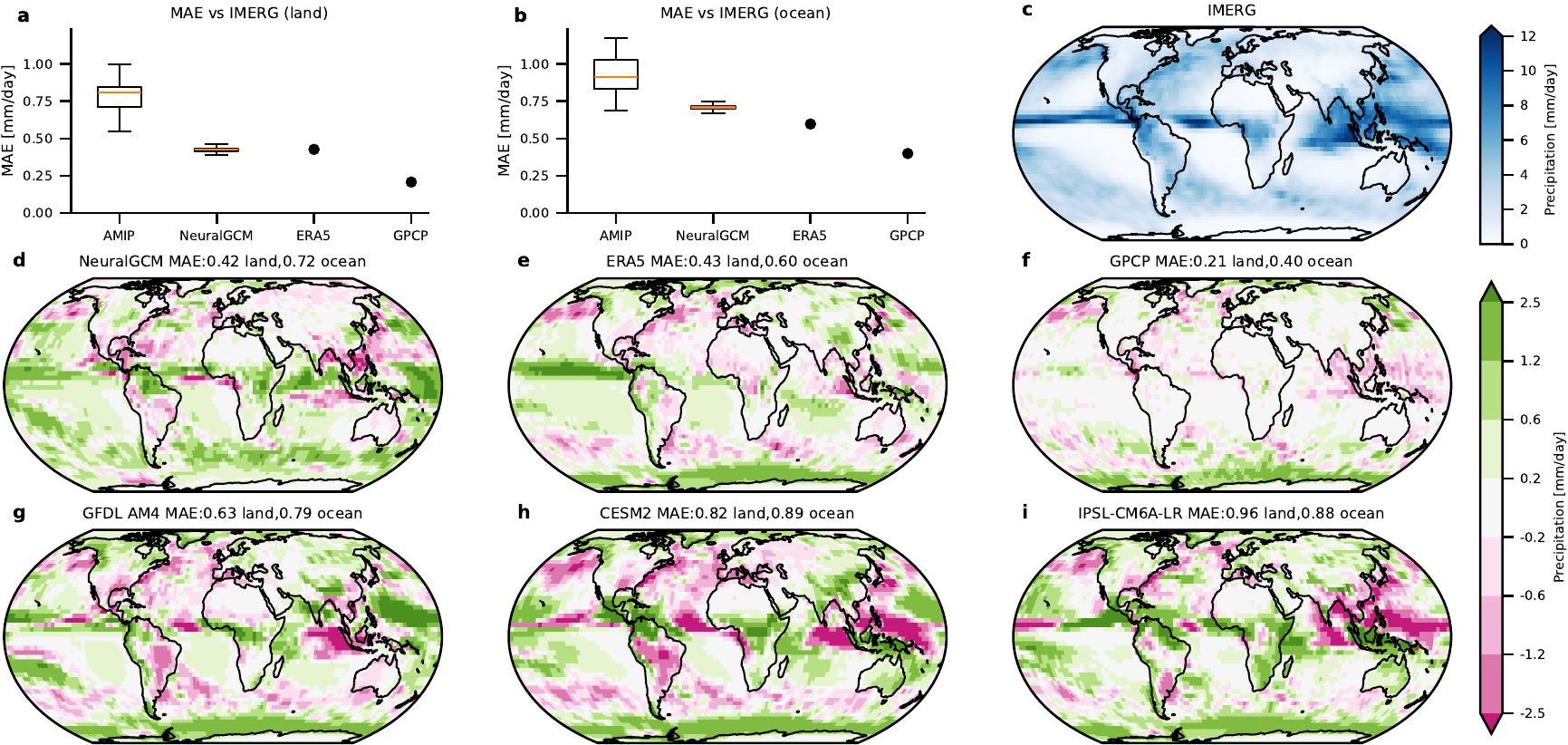}}}
\end{center}
\caption{
Mean bias in  precipitation averaged over 2002–2014 for September-October-November (same as Fig.~4 but for SON) 
}\label{sifig:mean_precip_SON}
\end{figure*}

\begin{figure*}
\begin{center}
\makebox[\textwidth]{\colorbox{white}{\includegraphics[width=0.65\paperwidth]{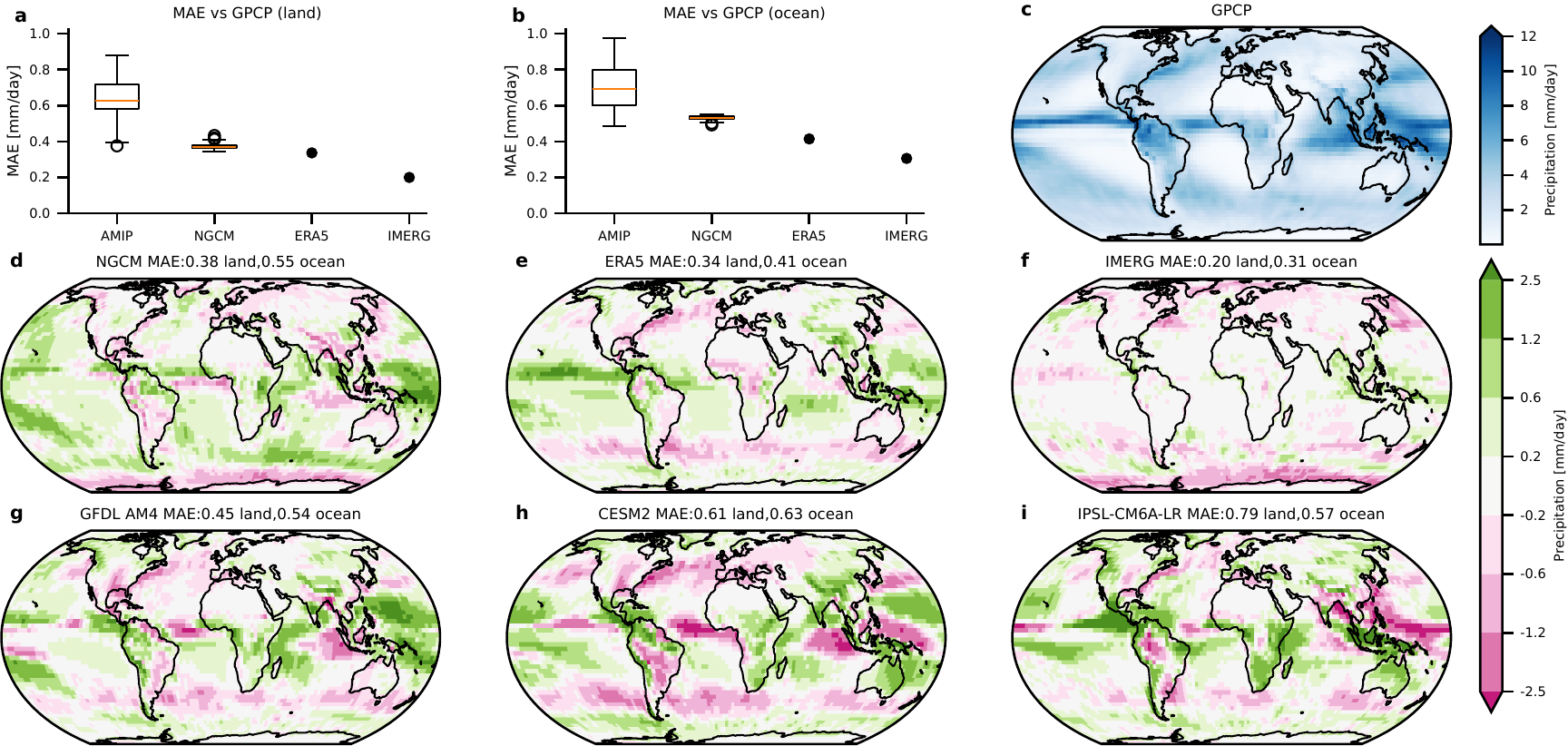}}}
\end{center}
\caption{
Mean bias in  precipitation averaged over 2002–2014 but using GPCP\cite{huffman2023new} as a baseline (same as Fig.~4 but using GPCP as a baseline) 
}\label{sifig:mean_precip_gpcp}
\end{figure*}

\begin{figure*}
\begin{center}
\makebox[\textwidth]{\colorbox{white}{\includegraphics[width=0.65\paperwidth]{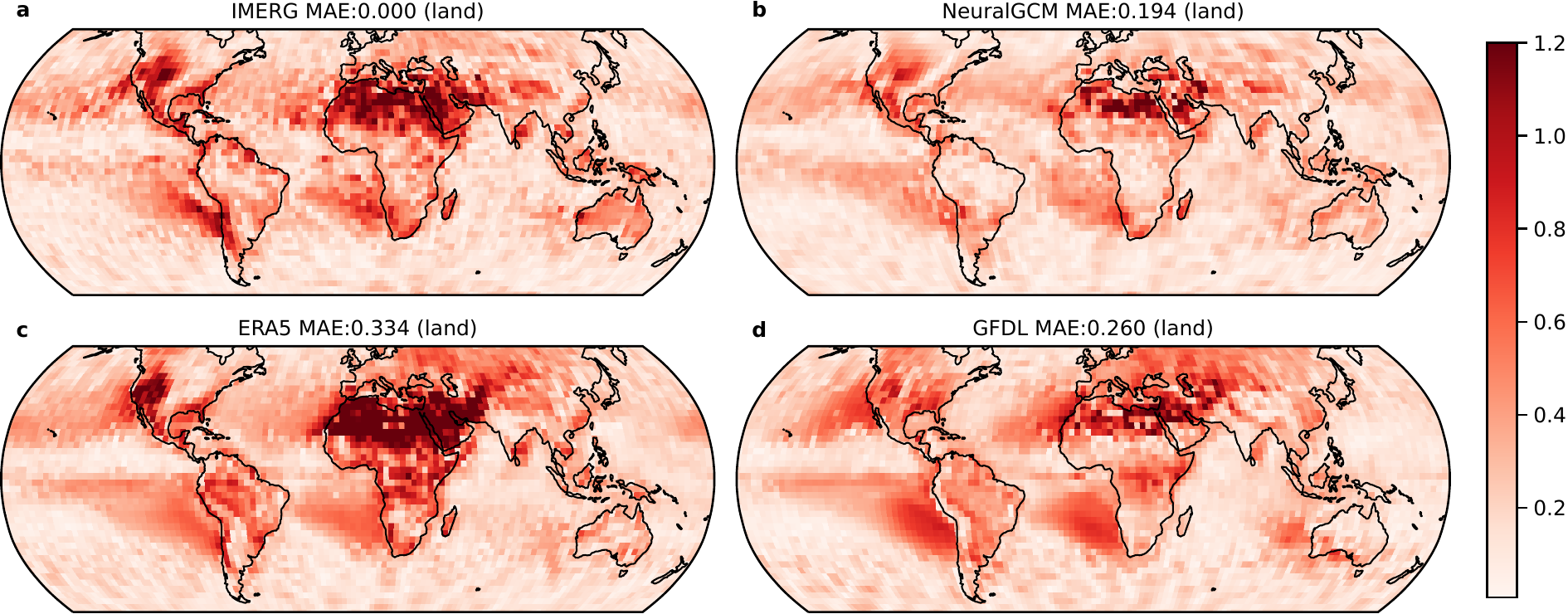}}}
\end{center}
\caption{
Diurnal harmonic amplitude of summertime precipitation (divided by the monthly mean precipitation; 2002-2014).
}\label{sifig:diurnal_amp}
\end{figure*}

\begin{figure*}
\begin{center}
\makebox[\textwidth]{\colorbox{white}{\includegraphics[width=0.65\paperwidth]{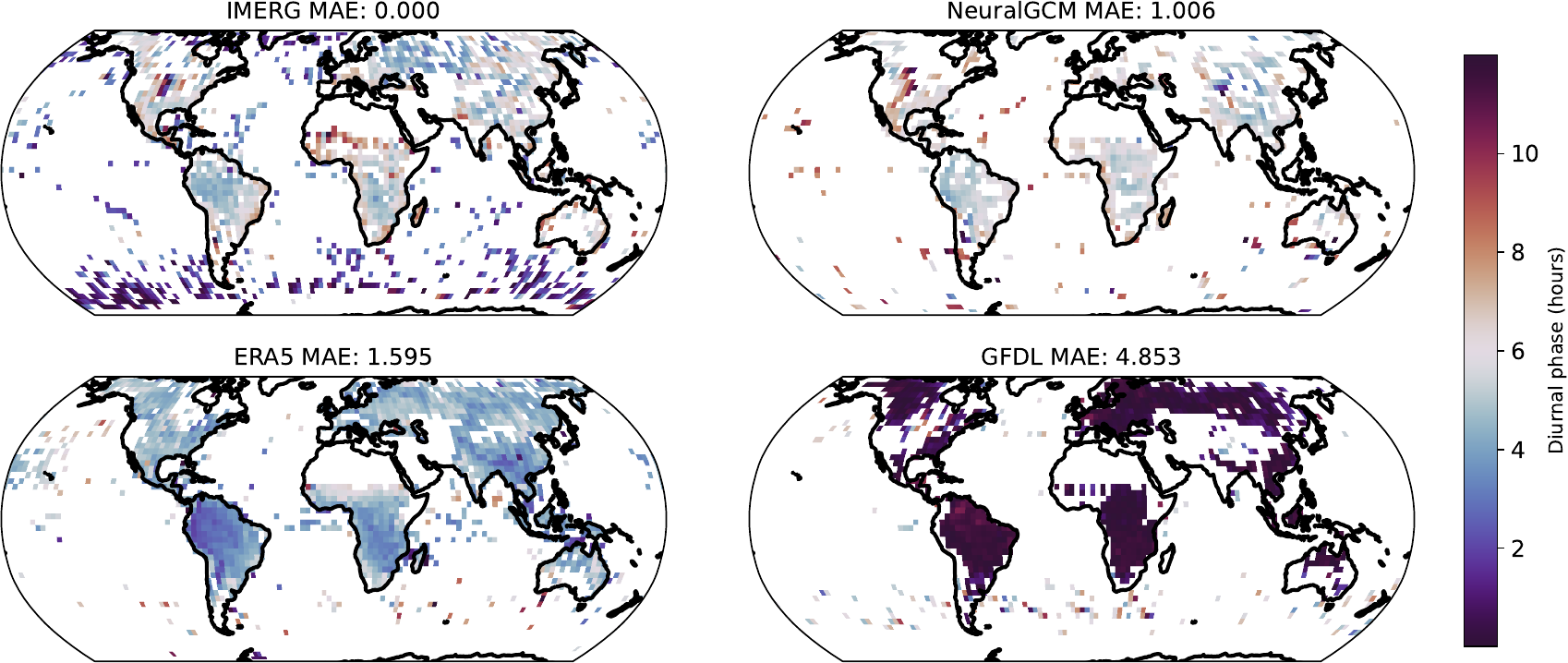}}}
\end{center}
\caption{
Semi-diurnal Cycle of Summertime Precipitation (2002-2014)
Same as Fig.~6 but for the semi-diurnal phase. 
}\label{sifig:semi_diurnal_cycle}
\end{figure*}

\begin{figure*}
\begin{center}
\makebox[\textwidth]{\colorbox{white}{\includegraphics[width=0.65\paperwidth]{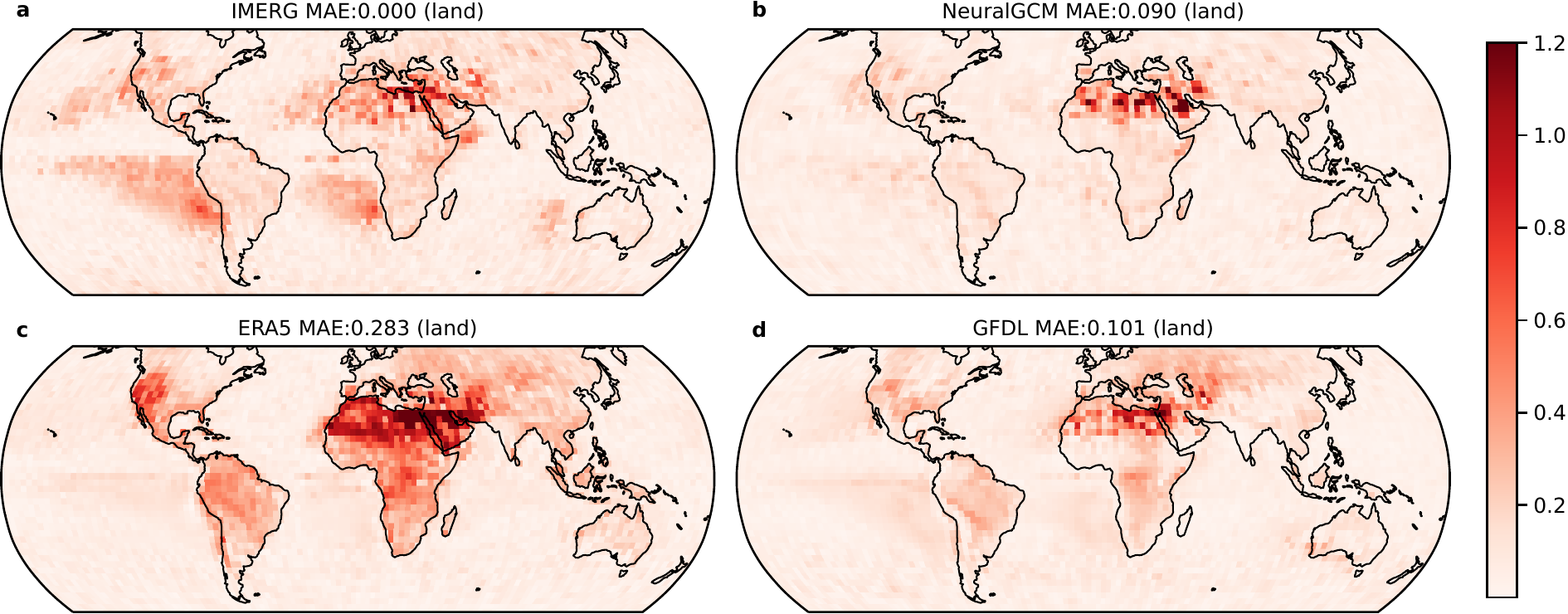}}}
\end{center}
\caption{
Semi-diurnal harmonic amplitude of summertime precipitation (divided by the monthly mean precipitation; 2002-2014).
}\label{sifig:semi_diurnal_amp}
\end{figure*}

\begin{figure*}
\begin{center}
\makebox[\textwidth]{\colorbox{white}{\includegraphics[width=0.45\paperwidth]{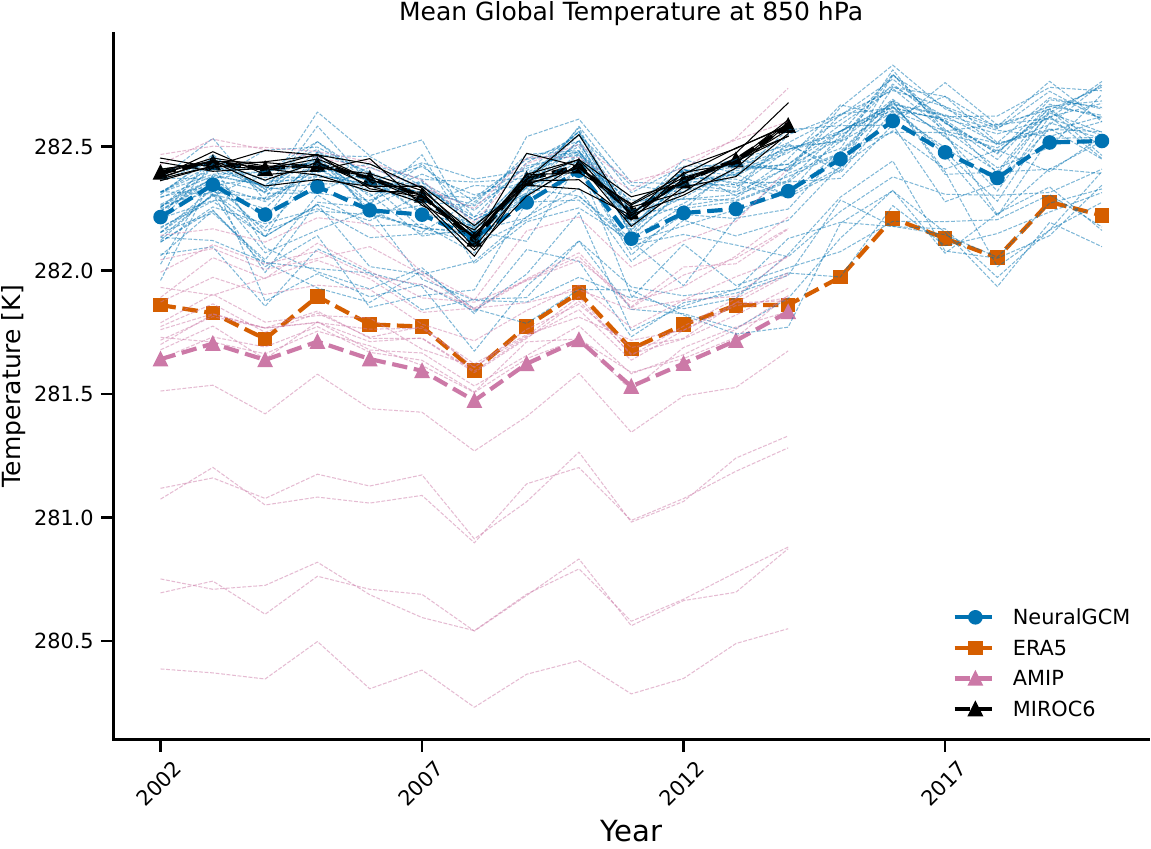}}}
\end{center}
\caption{
Global mean temperature for ERA5, NeuralGCM, CMIP6 AMIP runs, and 10 members of MIROC6 AMIP runs. 
Bold lines show the NeuralGCM ensemble mean, and AMIP models mean. 
AMIP models used in this plot are described in the methods.
}\label{sifig:global_mean_tempearture}
\end{figure*}

\begin{figure*}
\begin{center}
\makebox[\textwidth]{\colorbox{white}{\includegraphics[width=0.65\paperwidth]{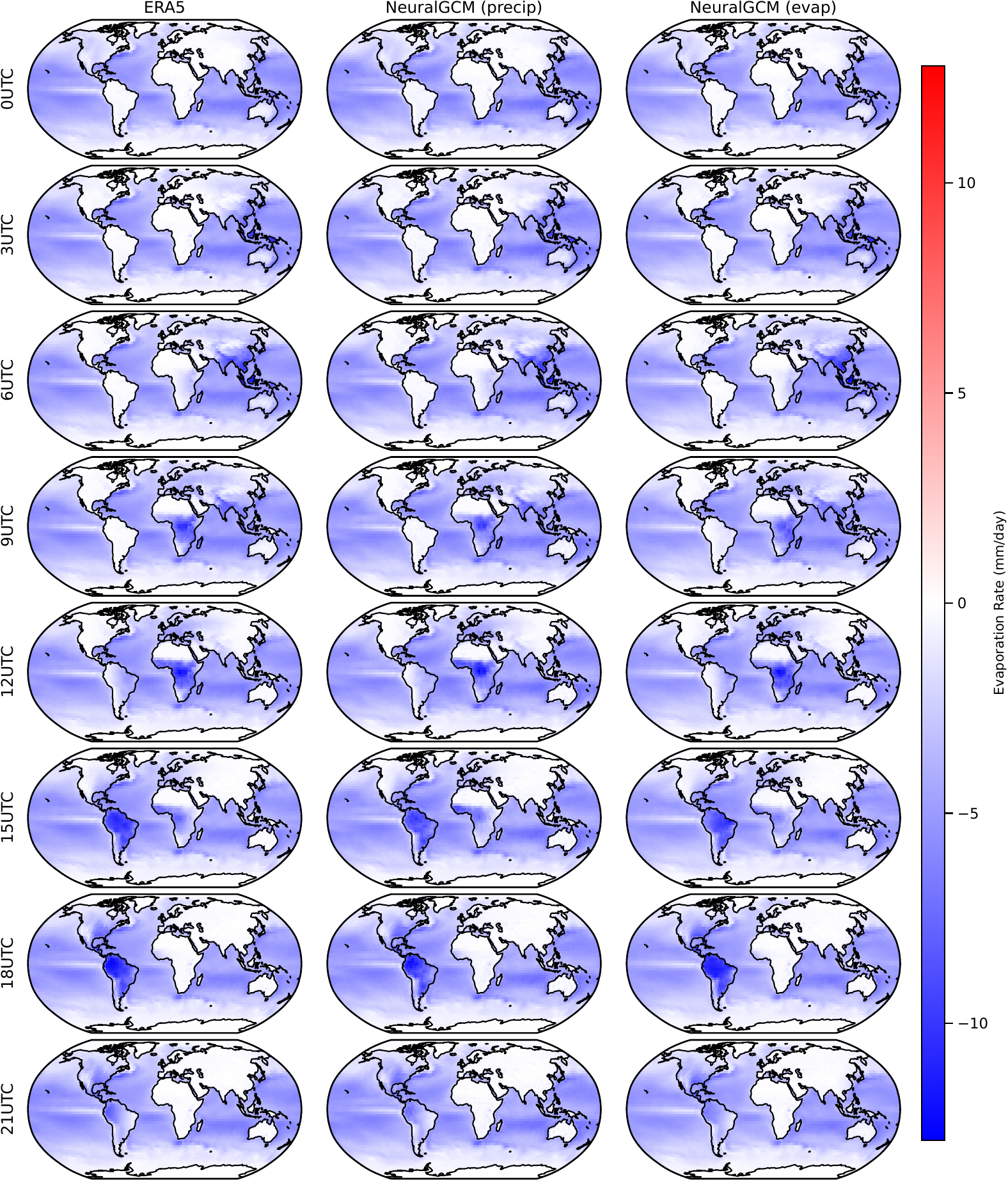}}}
\end{center}
\caption{
Diurnal cycle of hourly mean evaporation rate for ERA5 and two NeuralGCM configurations: (1) precipitation predicted, evaporation diagnosed; (2) evaporation predicted, precipitation diagnosed. Evaporation is averaged over 2020 (day 2 of the NeuralGCM simulations, initialized on December 27, 2001).
}\label{sifig:mean_evaporaiton}
\end{figure*}

\begin{figure*}
\begin{center}
\makebox[\textwidth]{\colorbox{white}{\includegraphics[width=0.65\paperwidth]{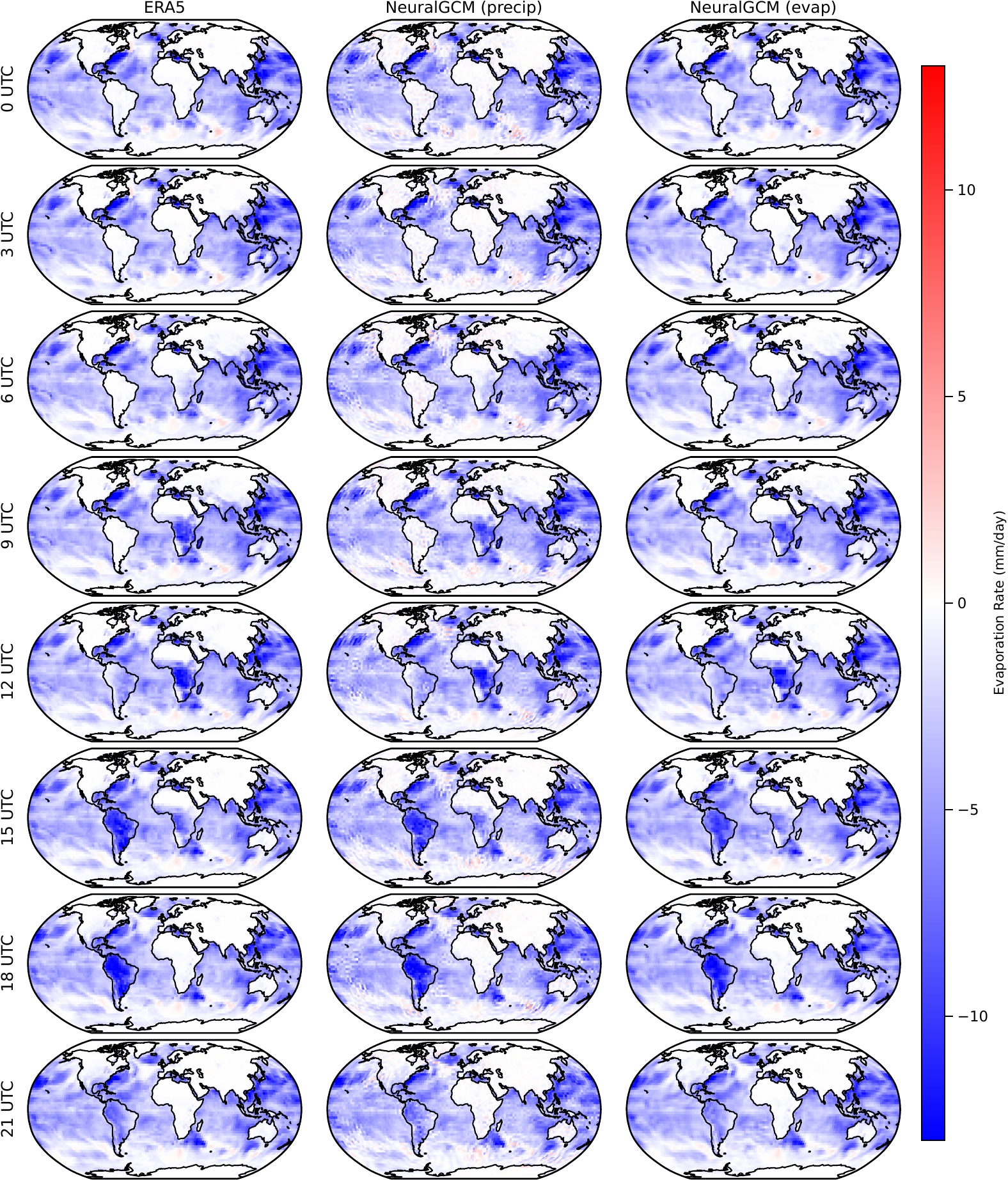}}}
\end{center}
\caption{
Diurnal cycle of instantaneous evaporation rate for ERA5 and two NeuralGCM configurations: (1) precipitation predicted, evaporation diagnosed; (2) evaporation predicted, precipitation diagnosed.  Shown for December 28, 2001 (day 2 of the NeuralGCM simulations, initialized on December 27, 2001).
}\label{sifig:instantenous_evaporaiton}
\end{figure*}


\begin{figure*}
\begin{center}
\makebox[\textwidth]{\colorbox{white}{\includegraphics[width=0.65\paperwidth]{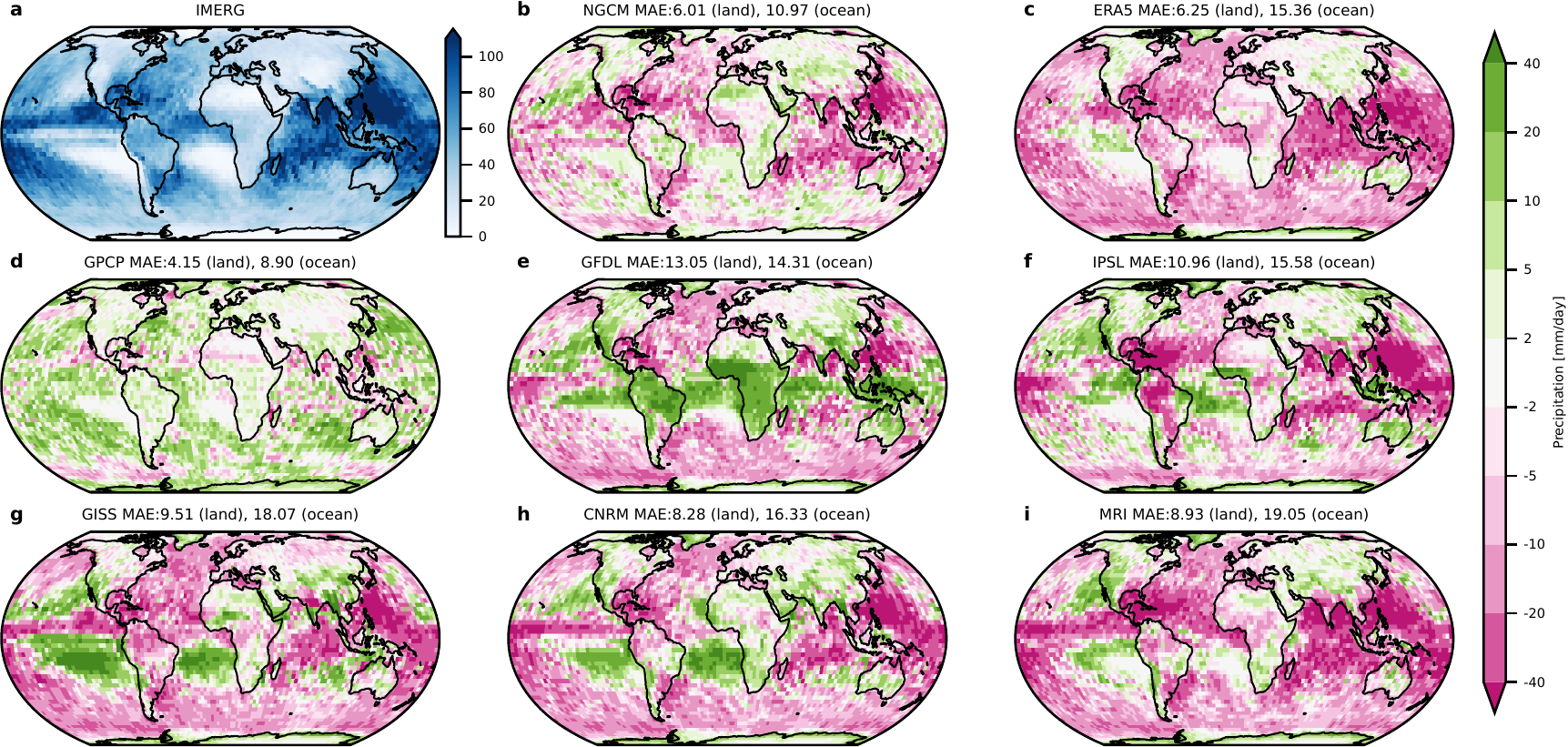}}}
\end{center}
\caption{
99.9th percentile of daily precipitation averaged over 2002–2014. (a) IMERG \cite{huffman2020integrated}. (b–i) Bias in the 99.9th percentile of precipitation for NeuralGCM, ERA5, GPCP\cite{huffman2023new}, and various CMIP6 historical simulations, relative to IMERG. Mean absoulute error (MAE) vs. IMERG is shown for land and ocean regions.
}\label{sifig:999th_precip}
\end{figure*}

\begin{figure*}
\begin{center}
\makebox[\textwidth]{\colorbox{white}{\includegraphics[width=0.65\paperwidth]{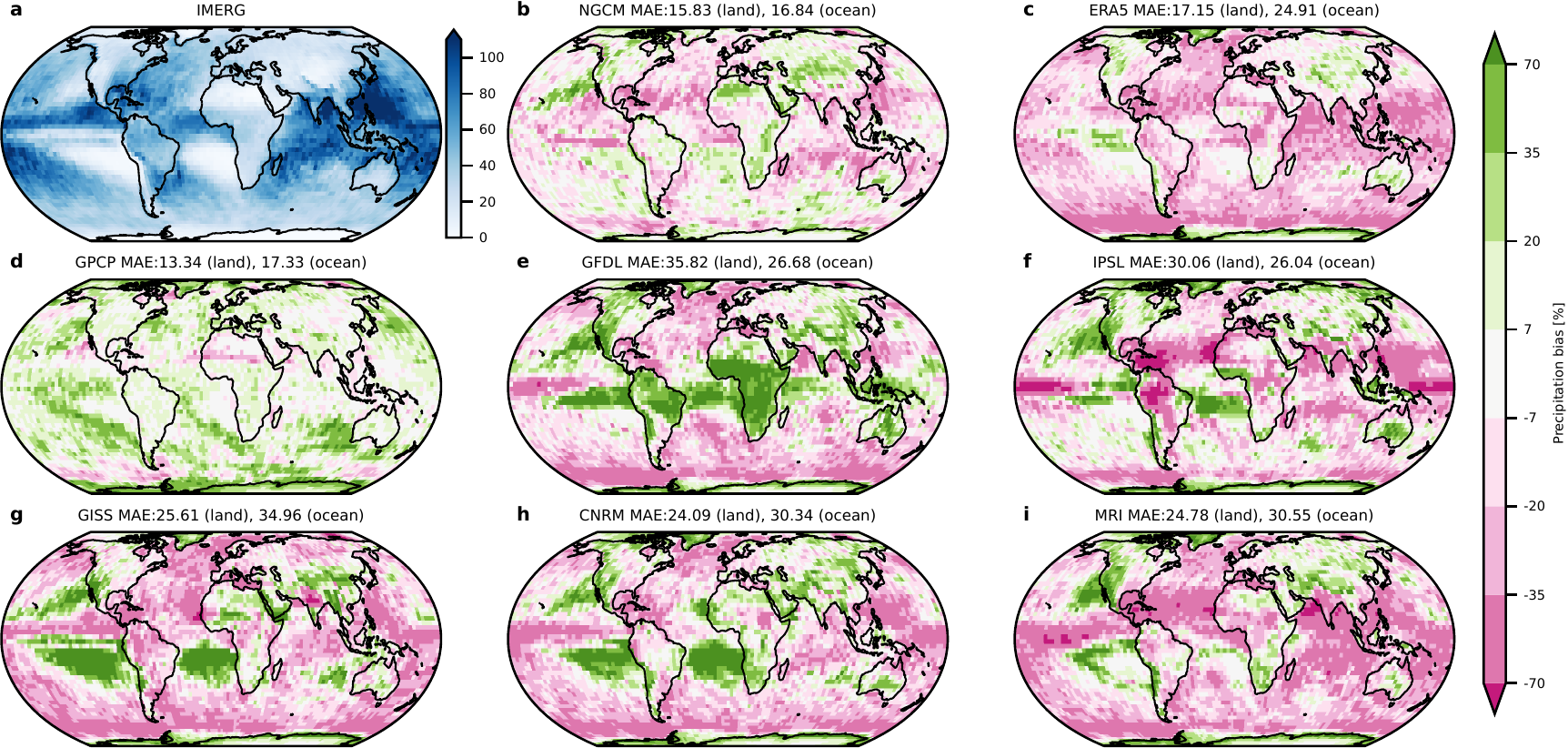}}}
\end{center}
\caption{
Percent mean absolute error in annual maximum daily precipitation (Rx1day) averaged over 2002–2014. (a) IMERG\cite{huffman2020integrated} in mm/day. (b–i) Percent error in Rx1day for NeuralGCM, ERA5, GPCP \cite{huffman2023new}, and various CMIP6 historical simulations, relative to IMERG. The percent error is calculated as  $(model - IMERG) / max(IMERG, 20)$ to de-emphasize errors in regions where Rx1day is lower than 20 mm/day. Global mean absolute error (in percent) is shown for land and ocean regions. 
}\label{sifig:rx1day_precip_percent}
\end{figure*}

\begin{figure*}
\begin{center}
\makebox[\textwidth]{\colorbox{white}{\includegraphics[width=0.65\paperwidth]{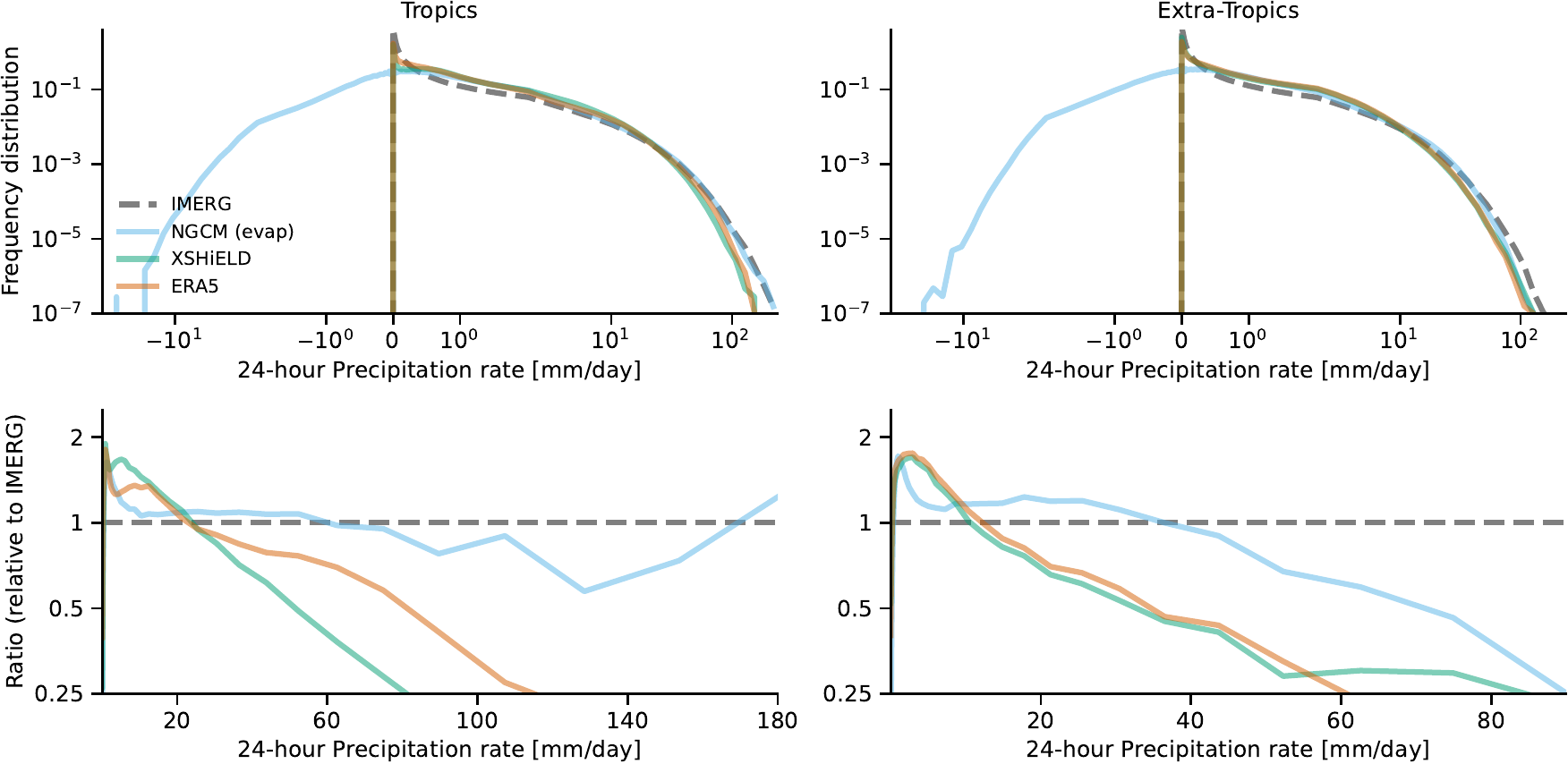}}}
\end{center}
\caption{Precipitation rate frequency distributions for IMERG, NeuralGCM (evap), ERA5 and X-SHEiLD. Like Fig.~\ref{sifig:distributions_precip_rate} but showing NeruralGCM-evap model that predicts evaporation and diagnose precipitation. The distribution shows that NeruralGCM-evap produces frequent negative precipitation. 
}\label{sifig:distributions_precip_rate_evap}
\end{figure*}

\begin{figure*}
\begin{center}
\makebox[\textwidth]{\colorbox{white}{\includegraphics[width=0.65\paperwidth]{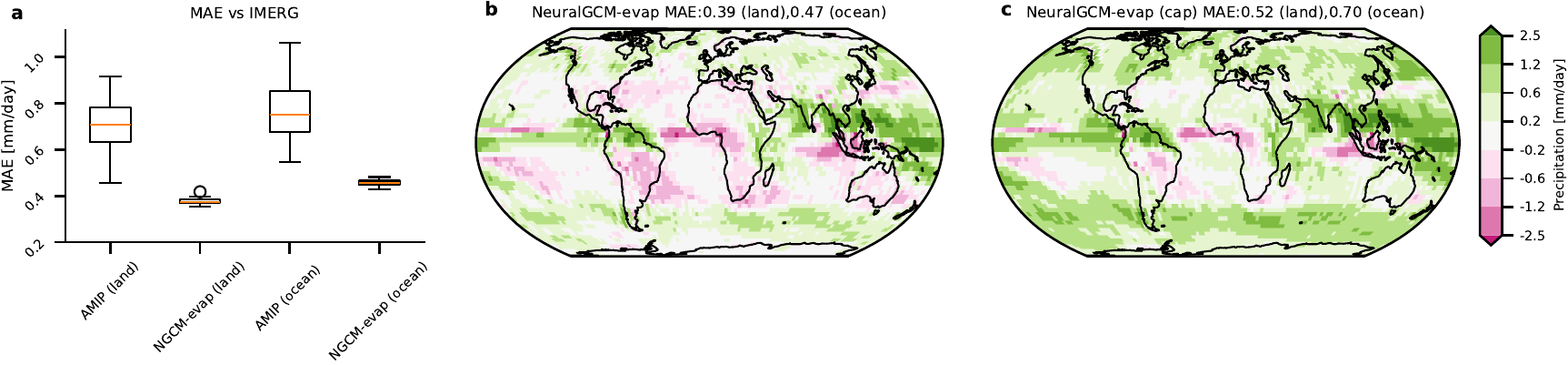}}}
\end{center}
\caption{
Mean absolute error (MAE) for precipitation averaged over 2002–2014 for the NeuralGCM-evap model. This figure is similar to Fig.~4 but shows results for: (a, b) the NeuralGCM-evap model, which predicts evaporation and diagnoses precipitation; and (c) the NeuralGCM-evap-cap model, a variant of NeuralGCM-evap where negative precipitation values are set to zero. Global MAE is shown for land and ocean regions.
}\label{sifig:mean_precip_evap}
\end{figure*}

\begin{figure*}
\begin{center}
\makebox[\textwidth]{\colorbox{white}{\includegraphics[width=0.65\paperwidth]{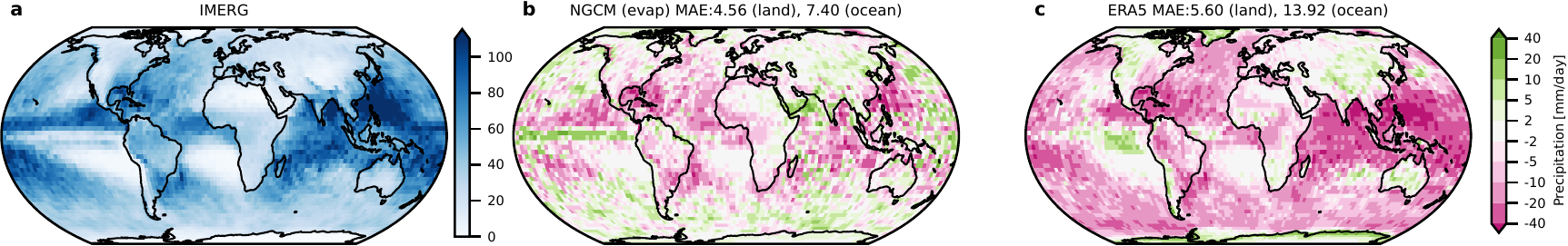}}}
\end{center}
\caption{Annual maximum daily precipitation (Rx1day) averaged over 2002–2014 for NeuralGCM-evap model. This figure is similar to Fig.~5c-e but shows results for (b) NeuralGCM-evap model which predicts evaporation and diagnose precipitation.
Global MAE is shown for land and ocean regions.
}\label{sifig:extreme_precip_evap}
\end{figure*}

\begin{figure*}
\begin{center}
\makebox[\textwidth]{\colorbox{white}{\includegraphics[width=0.65\paperwidth]{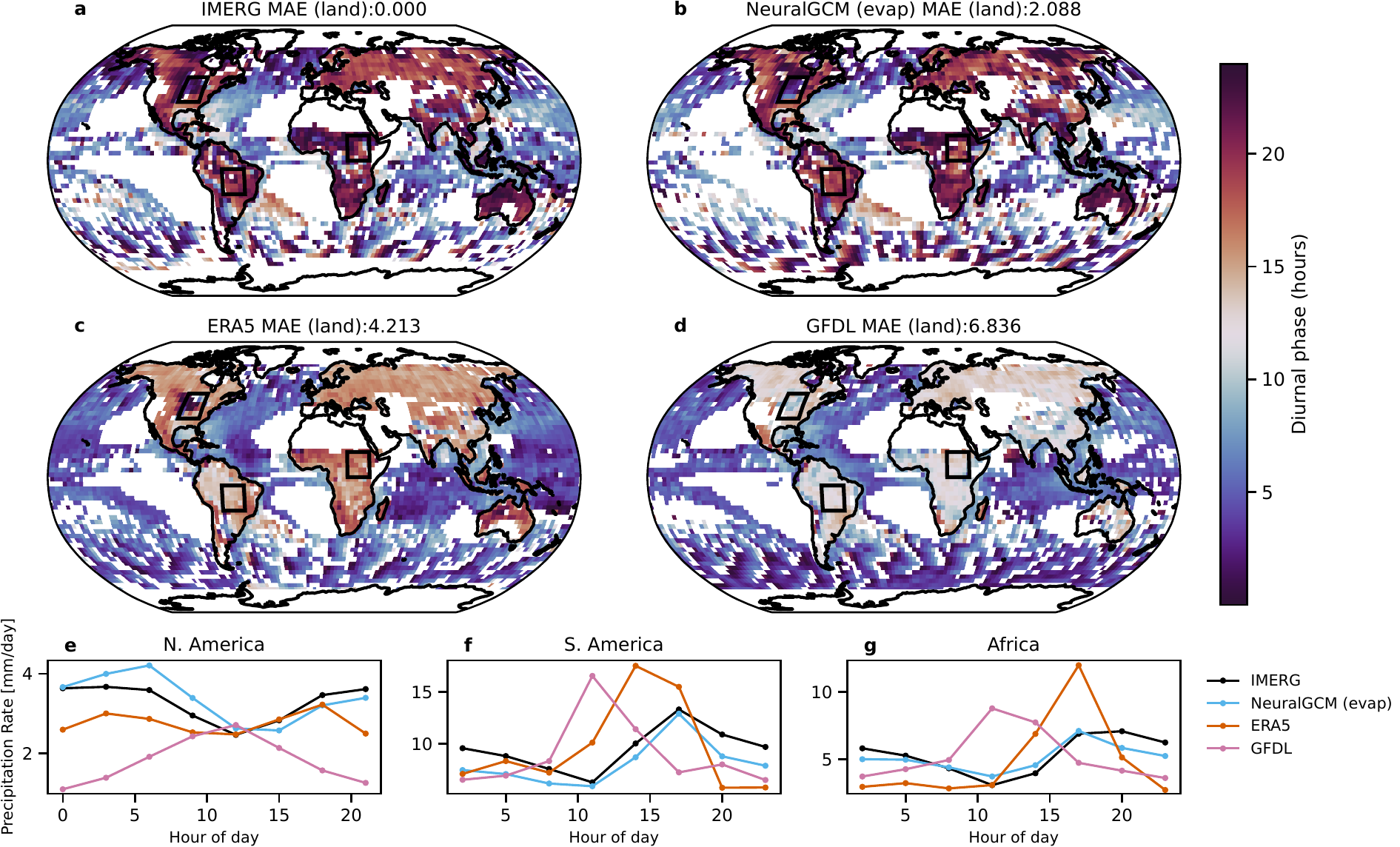}}}
\end{center}
\caption{
Diurnal Cycle of Summertime Precipitation (2002-2014) for NeuralGCM-evap.
Like Fig.~6 but showing results for NeuralGCM-evap model which predicts evaporation and diagnose precipitation.
}\label{sifig:diurnal_cycle_evap}
\end{figure*}

\begin{figure*}
\begin{center}
\makebox[\textwidth]{\colorbox{white}{\includegraphics[width=0.65\paperwidth]{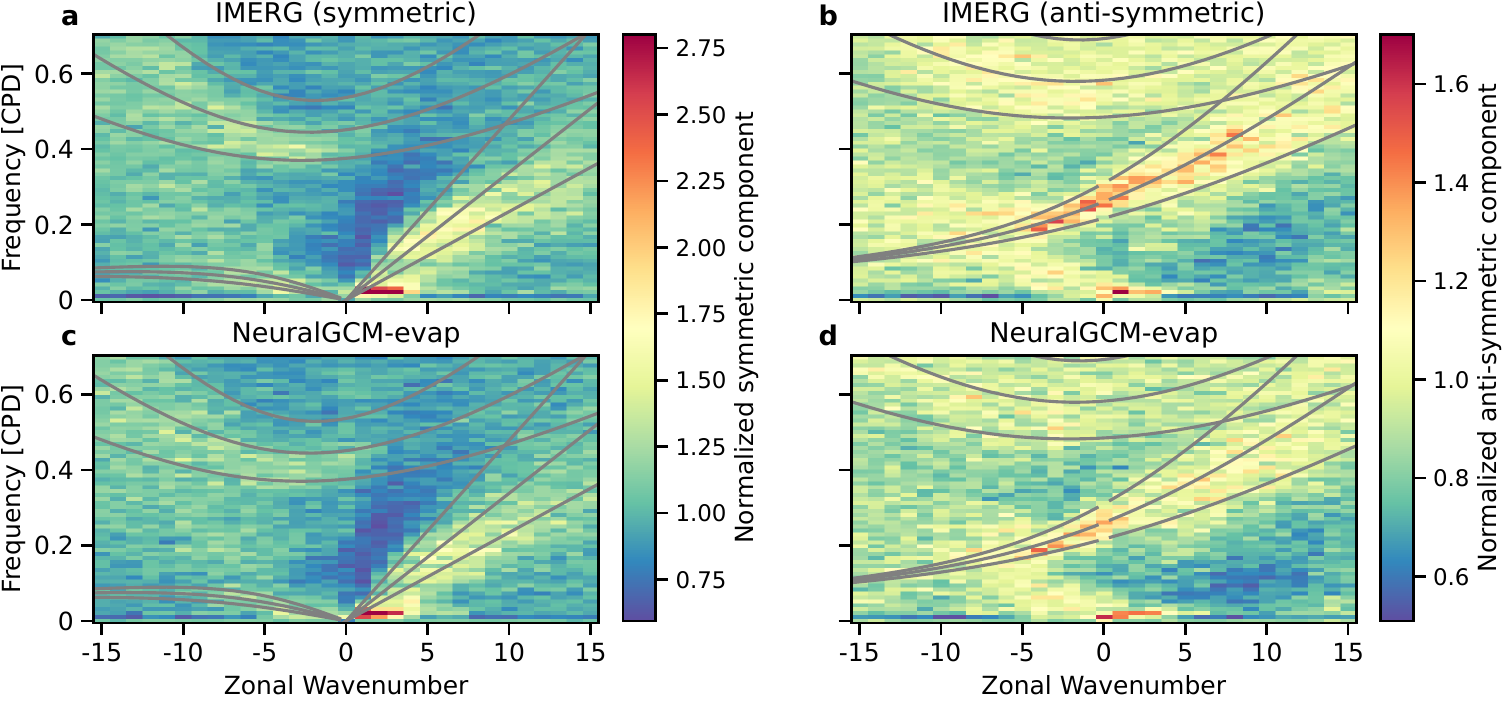}}}
\end{center}
\caption{
Space-Time Spectra of precipitation for IMERG and NeuralGCM-evap (2002-2014).
Like Fig.~\ref{sifig:Wheeler_Kiladis} but for NeuralGCM-evap model which predicts evaporation and diagnose precipitation.
}\label{sifig:Wheeler_Kiladis_evap}
\end{figure*}

\begin{figure*}
\begin{center}
\makebox[\textwidth]{\colorbox{white}{\includegraphics[width=0.45\paperwidth]{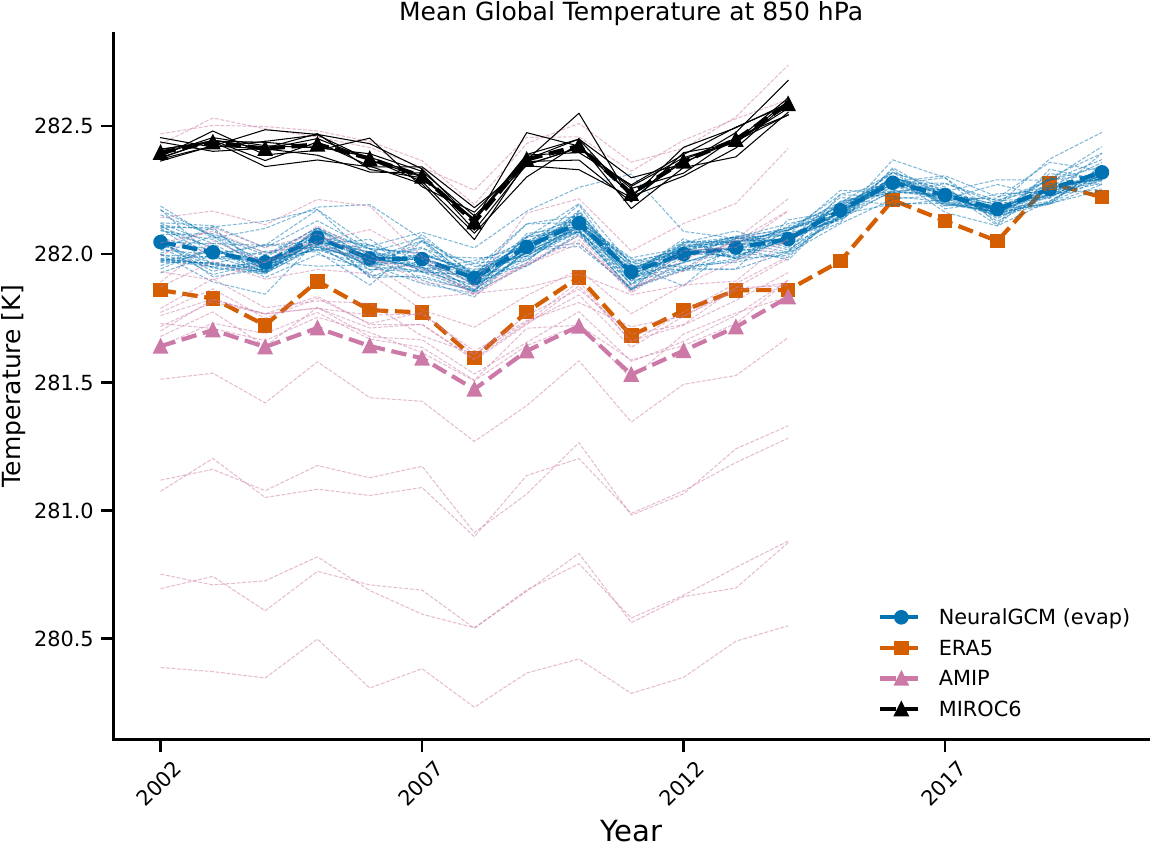}}}
\end{center}
\caption{
Global mean temperature for ERA5, NeuralGCM-evap, CMIP6 AMIP runs, and 10 members of MIROC6 AMIP runs. 
Same as Fig.~\ref{sifig:global_mean_tempearture}, but showing results for NeuralGCM-evap model, which predicts evaporation and diagnoses precipitation
}\label{sifig:global_mean_tempearture_evap}
\end{figure*}

\begin{figure*}
\begin{center}
\makebox[\textwidth]{\colorbox{white}{\includegraphics[width=0.6\paperwidth]{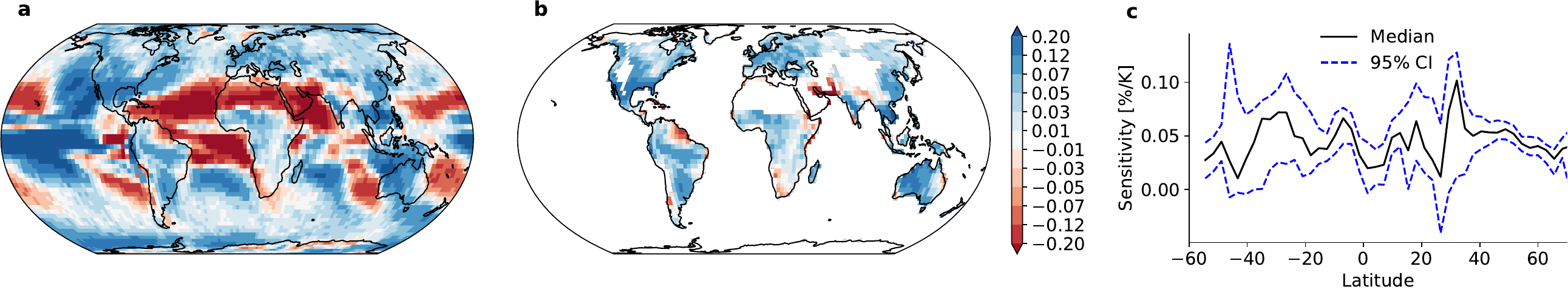}}}
\end{center}
\caption{
Sensitivity of annual-maximum daily precipitation (Rx1day) to changes in global mean temperature, calculated from NeuralGCM large ensemble runs over 2002–2021. (a) Sensitivity for all latitudes and longitudes, in units of \% K$^{-1}$. (b) Same as panel (a), but showing values only over land. (c) Median sensitivity over land (solid line; dotted lines show the 95\% confidence interval), averaged over longitudinal bands, as a function of global mean 850hPa temperature. Regions where Rx1day $<$ 20 mm/day where masked out. Sensitivity was derived using Theil-Sen regression of Rx1day against global mean temperature for each year and ensemble member in each grid box over the period 2002–2021. Confidence intervals were used from the Theil-Sen regression.
}\label{sifig:sensitivity_Rx1day}
\end{figure*}

\clearpage
\clearpage
\bibliographystyle{sn-nature.bst}
\bibliography{supplement}